%
\def\@{{\char'100}}

\long\def\abstract#1{\bigskip{\advance\leftskip by 2true cm
\advance\rightskip by 2true cm\eightpoint\centerline{\bf
Abstract}\everymath{\scriptstyle}\vskip10pt\vbox{#1}}\bigskip}
\long\def\resume#1{{\advance\leftskip by 2true cm
\advance\rightskip by 2true cm\eightpoint\centerline{\bf
R\'esum\'e}\everymath{\scriptstyle}\vskip10pt \vbox{#1}}}

\def\references{\bigbreak\centerline{\sc
References}\medskip\nobreak\bgroup
\def\ref##1&{\leavevmode\hangindent45pt
\hbox to 42pt{\hss\bf[##1]\ }\ignorespaces}
\parindent=0pt
\everypar={\ref}\par}
\def\endreferences{\egroup}
\long\def\authoraddr#1{\medskip{\baselineskip9pt\let\\=\cr
\halign{\line{\hfil{\Addressfont##}\hfil}\crcr#1\crcr}}}
\def\Subtitle#1{\medbreak\noindent{\Subtitlefont#1.} }
%
%
\newif\ifrunningheads
\runningheadstrue
\immediate\write16{- Page headers}
\headline={\ifrunningheads\ifnum\pageno=1\hfil\else\ifodd\pageno\rightheadline
\else\leftheadline\fi\fi\else\hfil\fi}
\def\rightheadline{\sc\hfil\RightHeadText\hfil}
\def\leftheadline{\sc\hfil\LeftHeadText\hfil}

\hyphenation{Harnad Neumann}
%
%
\immediate\write16{- Fonts "Small Caps" and "EulerFraktur"}
%
%
%

\let\sc=\tensmc
%
%
\font\teneuf=eufm10  \font\seveneuf=eufm7 \font\fiveeuf=eufm5
\newfam\euffam \def\gr{\fam\euffam\teneuf}

\textfont\euffam=\teneuf \scriptfont\euffam=\seveneuf
\scriptscriptfont\euffam=\fiveeuf
%

\def \wt {\widetilde}

\def \ra {\rightarrow}

\def \lra {\longrightarrow}

\def \a {\alpha}
\def \b {\beta}
\def \d {\delta}

\def \g {\gamma}
\def \G {\Gamma}
\def \k {\kappa}

\def \l {\lambda}

\def \th {\theta}
\def \t {\tau}
\def \o {\omega}

\def \ss {\subset}

\def \Lgl {\widetilde{\gr gl}}
\def \LGl {\widetilde{\gr Gl}}

\def \Lgl {\widetilde{\gr gl}}

\def \di {\partial}

\def \tr{{\rm tr}}
\def \det{{\rm det}}
\def \rank{{\rm rank}}
\def \diag{{\rm diag}}
\def\nchi{\hbox{\raise 2.5pt\hbox{$\chi$}}}
%
%

\def\grGl{{\gr Gl}}	\def\grgl{{\gr gl}}
\def\grSp{{\gr Sp}}	\def\grsp{{\gr sp}}

\def\nchi{\hbox{\raise 2.5pt\hbox{$\chi$}}}
%
%

\def\DD{{\cal D}}

\def\HH{{\cal H}}

\def\MM{{\cal M}}
\def\NN{{\cal N}}
\def\OO{{\cal O}}

%
%
		\def\bfA{{\bf A}}
		\def\bfB{{\bf B}}
		\def\bfC{{\bf C}}
		
\def\bfe{{\bf e}}		
\def\bff{{\bf f}}		\def\bfF{{\bf F}}
\def\bfg{{\bf g}}		\def\bfG{{\bf G}}
		
		\def\bfI{{\bf I}}
		
		\def\bfK{{\bf K}}
		
		\def\bfM{{\bf M}}

		\def\bfP{{\bf P}}
		\def\bfQ{{\bf Q}}
		\def\bfR{{\bf R}}
		\def\bfS{{\bf S}}
\def\bft{{\bf t}}		
		
\def\bfv{{\bf v}}

		\def\bfZ{{\bf Z}}

\def\tbe{\wt{\bfe}}
\def\tbf{\wt{\bff}}
\def\tbg{\wt{\bfg}}
\def\tbF{\wt{\bfF}}
\def\tbG{\wt{\bfG}}
\def\tH{\wt{H}}

\def\tth{\wt{\theta}}
\def\tchi{\wt{\chi}}
\def\tPsi{\wt{\Psi}}
\def\tK{\wt{K}}
\def\tbK{\wt{\bf{K}}}
\def\tR{\wt{R}}
\def\tbR{\wt{\bf R}}
\def\tbv{\wt{\bf v}}
%
%
\def\authorfont{\sc}
\font\eightrm=cmr8
\font\eightbf=cmbx8
\font\eightit=cmti8
\font\eightsl=cmsl8

\def\eightpoint{\let\rm=\eightrm \let\bf=\eightbf \let\it=\eightit
\let\sl=\eightsl \baselineskip = 9.5pt minus .75pt  \rm}

\font\titlefont=cmbx10 scaled\magstep2
\font\sectionfont=cmbx10
\font\Subtitlefont=cmbxsl10
\font\Addressfont=cmsl8
%
%
\def\Proclaim#1:#2\par{\smallbreak\noindent{\sc #1:\ }
{\sl #2}\par\smallbreak}
\def\Demo#1:#2\par{\smallbreak\noindent{\sl #1:\ }
{\rm #2}\par\smallbreak}
%
%
\immediate\write16{- Section headings}
\newcount\secount
\secount=0
\newcount\eqcount
\outer\def\section#1.#2\par{\global\eqcount=0\bigbreak
\ifcat#10
 \secount=#1\noindent{\sectionfont#1. #2}
\else
 \advance\secount by 1\noindent{\sectionfont\number\secount. #2}
\fi\par\nobreak\medskip}
%
%
\immediate\write16{- Automatic numbering}
\catcode`\@=11
\def\adv@nce{\global\advance\eqcount by 1}
\def\unadv@nce{\global\advance\eqcount by -1}
\def\nextnumber{\adv@nce}
%
%
\newif\iflines
\newif\ifm@resection
\def\onesec{\m@resectionfalse}
\def\moresec{\m@resectiontrue}
\moresec
\def\eq{\global\linesfalse\eq@}
\def\eqn{\global\linestrue&\eq@}
\def\nosubind@x{\global\subind@xfalse}
\def\newsubind@x{\ifsubind@x\unadv@nce\else\global\subind@xtrue\fi}
\newif\ifsubind@x
\def\eq@#1.#2.{\adv@nce
 \if\relax#2\relax
  \edef\loc@lnumber{\ifm@resection\number\secount.\fi
  \number\eqcount}
  \nosubind@x
 \else
  \newsubind@x
  \edef\loc@lnumber{\ifm@resection\number\secount.\fi
  \number\eqcount#2}
 \fi
 \if\relax#1\relax
 \else
  \expandafter\xdef\csname #1@\endcsname{{\rm(\loc@lnumber)}}
  \expandafter
  \gdef\csname #1\endcsname##1{\csname #1@\endcsname
  \ifcat##1a\relax\space
  \else
   \ifcat\noexpand##1\noexpand\relax\space
   \else
    \ifx##1$\space
    \else
     \if##1(\space
     \fi
    \fi
   \fi
  \fi##1}\relax
 \fi
 \eq@@{\loc@lnumber}}
\def\eq@@#1{\iflines \else \eqno\fi{\rm(#1)}}
\def\m@th{\mathsurround=0pt}
%
%
\def\display#1{\null\,\vcenter{\openup1\jot
\m@th
\ialign{\strut\hfil$\displaystyle{##}$\hfil\crcr#1\crcr}}
\,}
\newif\ifdt@p
\def\@lign{\tabskip=0pt\everycr={}}
\def\displ@y{\global\dt@ptrue \openup1 \jot \m@th
 \everycr{\noalign{\ifdt@p \global\dt@pfalse
  \vskip-\lineskiplimit \vskip\normallineskiplimit
  \else \penalty\interdisplaylinepenalty \fi}}}
%
%
\def\displayno#1{\displ@y \tabskip=\centering
 \halign to\displaywidth{\hfil$
\@lign\displaystyle{##}$\hfil\tabskip=\centering&
\hfil{$\@lign##$}\tabskip=0pt\crcr#1\crcr}}
%
%
\def\cite#1{{[#1]}}
\catcode`\@=\active
%
%
\magnification=\magstep1
\hsize= 6.75 true in
\vsize= 8.75 true in
%
%
\def\RightHeadText{Integrable Fredholm Operators and Dual Isomonodromic
Deformations}
\def\LeftHeadText{J. Harnad and A. Its}
%
\rightline{ CRM-2477 (1997) \break}
\rightline{solv-int/9706002 \break} \bigskip
\centerline{\titlefont Integrable Fredholm Operators and}
\centerline{\titlefont Dual Isomonodromic Deformations}
\bigskip
\centerline{\authorfont J.~Harnad}
\authoraddr
{Department of Mathematics and Statistics, Concordia University\\
7141 Sherbrooke W., Montr\'eal, Qu\'ebec, Canada H4B 1R6, {\rm \eightpoint
and} \\
Centre de recherches math\'ematiques, Universit\'e de Montr\'eal\\
C.~P.~6128, succ. centre ville, Montr\'eal, Qu\'ebec, Canada H3C 3J7\\
{\rm \eightpoint e-mail}: harnad\@crm.umontreal.ca}
\bigskip
\centerline{\authorfont  Alexander R.~Its}
\authoraddr
{Department of Mathematical Sciences, \\
Indiana University--Purdue University at Indianapolis\\
Indianapolis, IN 46202--3216
\\
{\rm \eightpoint e-mail}: itsa\@math.iupui.edu}
\bigskip
\abstract{The Fredholm determinants of a special class of integral
operators $\bfK$ supported on the union of $m$ curve segments in the complex
$\l$--plane are shown to be the $\t$-functions of an isomonodromic family of
meromorphic covariant derivative  operators $\DD_\l$. These have regular
singular points at the $2m$ endpoints of the curve segments and a singular
point of  Poincar\'e index $1$ at infinity. The rank $r$ of the vector bundle
over the Riemann sphere on which they act equals the number of distinct terms
in the exponential sums  entering in the numerator of the integral kernels.
The deformation equations may be viewed as nonautonomous Hamiltonian systems
on an auxiliary symplectic vector space $\MM$, whose Poisson quotient, under 
a parametric family of Hamiltonian group actions, is identified with a 
Poisson submanifold of the loop algebra  $\Lgl_R(r)$ with  respect to the
rational $R$--matrix structure. The matrix Riemann--Hilbert problem method is
used to identify the auxiliary space $\MM$ with the data defining the
integral kernel  of the resolvent operator at the endpoints of the  curve
segments.  A second associated isomonodromic family of covariant derivative
operators $\DD_z$ is derived,  having rank $n=2m$, and $r$ finite regular
singular points at the values of the exponents defining the kernel of $\bfK$.
This family is  similarly embedded into the  algebra $\Lgl_R(n)$ through a
{\it dual} parametric family of Poisson quotients of $\MM$. The operators 
$\DD_z$ are shown  to be analogously associated to the integral operator
$\wt{\bfK}$ obtained from $\bfK$ through a Fourier--Laplace transform.} 
\bigskip \baselineskip 14 pt

\section 1. Introduction
\medskip \nobreak

\Subtitle {1a. Dual Isomonodromic Deformation Equations}

\nobreak
The differential equations determining isomonodromic deformations of rational
covariant derivative operators of the form
$$
\eqalignno{
\DD_\l &={\di \over \di \l} - \NN(\l), \eqn DDl.a.\cr
\NN(\l)  &:= B + \sum_{j=1}^n {N_j \over \l -\a_j}, \eqn NNl.b.\cr
B &= \diag(\b_1, \dots  \b_r) , \quad N_j \in {\gr gl}(r,\bfC)}
$$
were derived in \cite{JMMS, JMU}. The residue matrices $N_j(\a_i, \b_a)$ may
depend parametrically on the pole locations $\{\a_1, \dots \a_n\}$ and the
asymptotic eigenvalues $\{\b_1, \dots \b_r\}$. This dependence must be such 
as to satisfy the commutativity conditions
$$
\left[\DD_\l, \ \DD_{\a_j}\right] =0, \quad
\left[\DD_\l, \ \DD_{\b_a}\right] =0, \quad
j=1, \dots n, \ a=1, \dots r,   \eq lamisomonintr..
$$
where the differential operators $\DD_{\a_j}, \ \DD_{\b_a}$ are defined by
$$
\eqalignno{
\DD_{\a_j} &:= {\di \over \di \a_j} + {N_j \over \l -\a_j}. \eqn.a. \cr
\DD_{\b_a}&:= {\di \over \di \b_{a}} - \l E_{a} -  \sum_{b=1 \atop b\neq a}^r
 {E_a \left(\sum_{j=1}^nN_j\right) E_b
+ E_b \left(\sum_{j=1}^nN_{j} \right)E_a \over \b_a - \b_b}, \eqn.b.}
$$
and $E_a$ is the elementary $r\times r$ matrix with elements
$$
(E_a)_{bc}:= \d_{ab} \d_{ac}. \eq Eaintr..
$$

  In the Hamiltonian formulation \cite{H} these relations, viewed as
differential  equations for the residue matrices $N_i$, are interpreted as a
compatible system of nonautonomous Hamiltonian equations with respect to the
Lie Poisson structure on the space $\left({\gr gl}(r)\right)^{*n} = \{N_1,
\dots N_n\}$, embedded through formula \NNl into the loop algebra $\Lgl_R(r)$
with rational $R$--matrix structure.  They are generated by the Poisson
commuting family of Hamiltonians $\{H_j, \ K_a\}_{j=1 \dots n, a=1\dots r}$
defined by $$
\eqalignno{
H_j &:=\tr(BN_{j}) + \sum_{i=1 \atop i\neq
j}^{n}{\tr\left(N_{i}N_{j}\right)\over
 \a_{j}-\a_{i}}
\eqn Hkintr.a. \cr
K_a &:=
 \sum_{j=1}^{n}\a_{j}\left(N_{j}\right)_{aa} + \sum_{b=1 \atop b\neq a}^r
{{\left(\sum_{j=1}^{n}N_{j}\right)_{ab}\left(\sum_{k=1}^{n}N_{k}\right)_{ba}}
\over \b_{a}-\b_{b}}. \eqn Kaintr.b.}
$$
The Poisson commutativity of the Hamiltonians follows from the $R$--matrix
structure on $\Lgl_R(r)$ and implies that the differential $1$--form on the
parameter space $\{\a_, \dots \a_n, \b_1, \dots \b_r\}$ defined by
$$
 \o :=\sum_{k=1}^n H_k d\a_k + \sum_{a=1}^r K_ad\b_a  \eq omegintr..
$$
is closed
$$
d\o=0.  \eq omegclosed..
$$
This, in turn, implies the existence of a tau-function
$\t(\a_1, \dots \a_n; \b_1, \dots \b_r)$ in the sense of \cite{JMU}, 
determined, up to multiplication by a local constant, by
$$
\o = d(\ln \t).  \eq tauintr..
$$

   As shown in \cite{H} it is natural to view the phase space as
a Poisson quotient of a  symplectic space $\MM=\{F, G\}$
consisting of canonically conjugate pairs $(F,G)$ of rectangular $N\times r$
matrices, where
$$
N = \sum_{i=1}^n k_i, \quad k_i:= \rank{N_i}.  \eq..
$$
The quotient map is defined by the formula
$$
\NN(\l)= B+F^T(A-\l\bfI)^{-1}G,  \eq NNintr..
$$
where $A\in {\gr gl}(N, \bfC)$ is a diagonal matrix with eigenvalues
$\{\a_i\}_{i =1 \dots n}$ and multiplicities $\{k_i\}$. The Hamiltonians may
be pulled back to $\MM$ through the projection map defined by \NNintr, and 
hence determine corresponding nonautonomous Hamiltonian systems
$$
\eqalignno{
{\di F \over \di \a_i} &= \{F, \ H_i\}, \quad
{\di G \over \di \a_i} = \{G, \ H_i\}, \quad i=1 \dots n, 
\eqn FHameqintr.a.\cr
{\di F \over \di \b_a} &= \{F, \ K_a\}, \quad
{\di G \over \di \b_a} = \{G, \ K_a\} \quad  a= 1, \dots r,
\eqn GHameqintr.b.\cr }
$$
\nextnumber
whose integral curves project to solutions of the isomonodromic deformation
equations \lamisomonintr.

   This leads quite naturally to a ``dual'' isomonodromic deformation system 
for the family of operators
$$
\eqalignno{
\wt{\DD}_z &:= {\di \over \di z} -\MM(z),  \eqn DDzintr.a. \cr
\MM(z)&:= A+F(B-z\bfI)^{-1}G^T = A + \sum_{a=1}^r {M_a \over z -\b_a},
\eqn MMintr.b.\cr
M_a &\in {\gr gl}(N, \bfC).}
$$
The corresponding deformations equations are
$$
\left[\wt{\DD}_z,\ \wt{\DD}_{\a_j}\right]=0, \quad
\left[\wt{\DD}_z,\ \wt{\DD}_{\b_a}\right], \quad
j=1,\dots n, \quad a=1, \dots r,   \eq zisomonintr..
$$
where
$$
\eqalignno{
\wt{\DD}_{\b_a} &:={\di \over \di \b_a} + {M_a\over z-\b_a}, 
\eqn Datintr.a. \cr
\wt{\DD}_{\a_i}&:= {\di \over \di \a_{i}} - z E_{i} - 
\sum_{j=1 \atop j\neq i}^n {E_i \left(\sum_{a=1}^r M_a\right) E_j
+ E_j \left(\sum_{a=1}^r M_{a}\right) E_i \over \a_i - \a_j}.
\eqn Ditintr.b.}
$$
These have the same interpretation, as nonautonomous Hamiltonian equations
on the phase space $\left({\gr gl}(n)\right)^{*r} = \{M_1, \dots M_r\}$,
generated by the Poisson commuting family  of Hamiltonians
$$
\eqalignno{
\tK_a &:=\tr(A M_{a}) + \sum_{b=1 \atop b\neq a}^{r}
{\tr \left(M_{a}M_{b}\right)\over
 {\b_{a}-\b_{b}}},   \eqn tHkintr.a. \cr
\tH_j &:= \sum_{a=1}^{r}\b_{a}\left(M_{a}\right)_{jj} + 
\sum_{k=1 \atop k\neq
j}^n { {\left (\sum_{a=1}^{r}M_{a}\right )_{jk}
\left (\sum_{b=1}^{r}M_{b}\right)_{kj}} \over{\a_{j}-\a_{k}}}.
\eqn tKaintr.b.}
$$
\nextnumber
Pulled back to the space $\MM$, these are shown in \cite{H} to coincide with
the Hamiltonians \Hkintr, \Kaintr, and hence the corresponding ``dual''
$\t$--function $\wt{\t}(\a_1, \dots \a_n; \b_1 \dots \b_r)$,  determined by
$$
d(\ln \wt{\t}) = :=\sum_{k=1}^n \tH_k d\a_k + \sum_{a=1}^r \tK_a d\b_a,
\eq dualtauintr..
$$
coincides with $\t(\a_1, \dots \a_n; \b_1 \dots \b_r)$.

   In the present work, using the approach of \cite{IIKS}, it will be
shown  how a particular class of solutions to such isomonodromic deformation
equations arises through the solution of an associated matrix 
Riemann--Hilbert problem (Theorems 2.1, 2.3), and how the corresponding
$\t$--functions may be identified with the Fredholm determinants of a simple
class of ``integrable'' Fredholm integral operators (Theorem 2.5).  We also
show (Theorem 2.6) that the $\t$--function obtained in this way may
equivalently be interpreted as a $\t$--function in the sense of Sato
\cite{Sa} and Segal and Wilson \cite{SW}; that is, as the determinant of a
suitably defined projection operator applied to the image of an element $W$
of an infinite dimensional Grassmann manifold  under the action of an abelian
subgroup $\G^r\ss \LGl(r)$ of the loop group $\LGl(r)$. The ``dual'' system
is then shown (Theorem 3.1), under suitable restrictions on the input data,
to correspond to the integral operator obtained through a Fourier--Laplace
transform.  In this formulation, the element $W$ of the Grassmannian is
determined by the pole parameters $\{\a_1, \dots \a_n\}$ of the first family
of operators \NNl, while the group element  $\g \in \G^r$ is parameterized by
the asymptotic eigenvalues $\{\b_1, \dots \b_r\}$. In the dual system, the
r\^oles of the element $W$ and the group element $\g$ of are interchanged.

\medskip
\Subtitle {1b. Integrable Fredholm Kernels and the Riemann--Hilbert Problem}

\nobreak
Consider a $p \times p$ matrix Fredholm integral operator acting on
${\bf C}^p$--valued functions ${\bf v}(\l)$,
$$
\bfK ({\bfv})(\l) = \int_\G K(\l, \mu) {\bfv}(\mu)d\mu, \eq FredK..
$$
defined along a piecewise smooth, oriented curve $\G$ in the complex plane
(possibly extending to $\infty$), with integral kernel of the special form
$$
K(\l, \mu) = {\bff^T(\l)\bfg(\mu)\over \l - \mu},  \eq Kdef..
$$
where $\bff, \bfg$ are rectangular $r\times p$ matrix valued functions,
$p\le r$.
For the moment, we assume only that $\bff$ and $\bfg$ are smooth functions
along
the connected components of $\G$ and, in order that $\bfK$ be nonsingular, we
also  require  that the numerator vanish at least linearly in $\l-\mu$ in the
limit  $\l\ra \mu$,
$$
\bff^T(\l)\bfg(\l) =0,  \eq fgnull..
$$
so that the diagonal values are given by taking the limit
$$
K(\l,\l)= \bff^{\prime T}(\l) \bfg(\l)= - \bff^T(\l)\bfg^{\prime}(\l).
\eq Kdiag..
$$

  Fredholm determinants of operators of this type appear as generating
functions  for correlators in many integrable quantum field theory models
\cite{WMTB, JMMS, IIKS, IIKV, KBI} and as spectral distributions for random
matrix  ensembles  \cite{M, TW1--2,  HTW}, the most common case being $p=1$,
$r=2$, with  ${\bf K}$ a scalar integral operator. A first important
observation to note \cite{IIKS} is that the resolvent operator
$$
\bfR := (\bfI-\bfK)^{-1}\bfK  \eq Resolv..
$$
is also in the same class. Thus, $\bfR$ may also be expressed as
$$
\bfR (\bfv)(\l) = \int_\G R(\l, \mu) \bfv(\mu)d\mu \eq..
$$
where the resolvent kernel
$$
R(\l, \mu) :=  {\bfF^T(\l)\bfG(\mu)\over \l - \mu}.  \eq..
$$
is determined by
$$
\eqalignno{
\bfF^T &= (\bfI-\bfK)^{-1} \bff^T = (\bfI  + \bfR)\bff^T \eqn FfK.a.  \cr
\bfG &= \bfg(\bfI-\bfK)^{-1} =\bfg(\bfI  + \bfR), \eqn GgK.b.}
$$
\nextnumber
where  $(1-\bfK)^{-1}$ in \FfK is understood as acting to the right, while
in \GgK it acts to the left.
These quantities similarly satisfy the nonsingularity condition
$$
\bfF^T(\l)\bfG(\l)=0,  \eq FGnull..
$$
with the diagonal value of the resolvent kernel given by
$$
R(\l,\l) = \bfF^{T \prime}(\l) \bfG(\l) =  - \bfF^{T}(\l) \bfG^{\prime}(\l).
\eq Rdiag..
$$
The basic deformation formula relating the Fredholm determinant to the
resolvent operator is
$$
d\ln \det(\bfI- \bfK) = -\tr\left( (1-\bfK)^{-1}d\bfK \right)
= -\tr \left( (1+\bfR)d\bfK \right),
\eq Freddetdiff..
$$
where $d$ signifies the differential with respect to any auxiliary parameters
on which $\bfK$ may depend.

  The determination of $\bfR$ is equivalent to the solution of an
associated matrix Riemann--Hilbert problem (cf. \cite{IIKS}). Define the
$r\times r$ invertible matrix $H(\l)$ as the following  rank-$p$ perturbation
of the identity matrix $\bfI_r$
$$
H(\l) = \bfI_r +2\pi i \bff(\l)\bfg^T(\l) = \exp 2\pi i \bff(\l)\bfg^T(\l)
\eq Hl..
$$
(where the exponential form holds here because the nonsingularity
condition \fgnull implies that $\bff(\l) \bfg^T(\l)$ is nilpotent.)
 The relevant matrix Riemann--Hilbert problem consists of
finding a nonsingular $r\times r$ matrix valued function $\chi (\l )$ that is
analytic on the  complement of $\G$, extends to $\l=\infty$ off $\G$, with
asymptotic form
$$
\chi(\l) \sim \bfI_r + \OO\left(\l^{-1}\right)  \eq MRHasympt..
$$
for $\l \ra \infty$, and has cut discontinuities across $\G$ given by
$$
\chi_-(\l) = \chi_+(\l) H(\l), \quad \l \in \G,  \eq MRHdisc..
$$
where $\chi_+(\l)$ and $\chi_-(\l)$ are the limiting values of $\chi(\l)$ as
$\G$  is approached from the left and the right, respectively.  From the 
Cauchy integral representation we then have
$$
\eqalignno{
\chi(\l) &= \bfI_r - \int_\G {\chi(\mu)\bff(\mu) \bfg^T(\mu)
\over \mu -\l} d\mu,  \eqn chiint.a. \cr
\chi^{-1}(\l) &= \bfI_r + \int_\G { \bff(\mu)\bfg^T(\mu) \chi^{-1}(\mu)\over
\mu -\l} d\mu, \eqn chiinvint.b. }
$$
\nextnumber
where the choice of branch $\chi_{\pm}(\mu)$ in the integrand is immaterial, 
in view of \fgnull. It  follows that the $p\times r$ matrix--valued functions
$\bfF(\l), \ \bfG(\l)$ defining the resolvent kernel are obtained by applying
the  matrices $\chi(\l)$, $(\chi^T)^{-1}(\l)$ to  $\bff(\l)$ and $\bfg(\l)$,
respectively
$$
\eqalignno{
\bfF(\l) &= \chi(\l)\bff(\l), \eqn Fdr.a. \cr
\bfG(\l) &= (\chi^T)^{-1}(\l)\bfg(\l).    \eqn Gdr.b.}
$$
\nextnumber
The matrices $\chi(\l)$ , $\chi^{-1}(\l)$ are thus determined on the 
complement of $\G$ by the integral formulae
$$
\eqalignno{
\chi(\l) &=  \bfI_r + \int_\G {\bfF(\mu)\bfg^T(\mu) \over \l -\mu} d\mu,
\eqn chiaab.a. \cr
\chi^{-1}(\l) &=  \bfI_r - \int_\G {\bff(\mu) \bfG^T(\mu) \over \l - \mu}
d\mu, \eqn chiinvaab.b.}
$$
\nextnumber
with appropriate limiting values as $\l\ra \G$.
It follows that if we express the asymptotic form of $\chi(\l)$ and
$\chi^{-1}(\l)$ for large $\l$ as
$$
\eqalignno{
\chi(\l) &\ {\sim }\ \bfI_r +\sum_{j=1}^\infty{\chi_j\over \l^j},
\eqn chiasympt.a. \cr
\chi^{-1}(\l) &\ {\sim }\ \bfI_r +\sum_{j=1}^\infty{\hat{\chi}_j\over \l^j},
\eqn chiinvasympt.b.}
$$
\nextnumber
the aymptotic coefficients $\chi_j, \ \hat{\chi_j}$ are given by
$$
\eqalignno{
\chi_j &= \int_{\G} \l^{j-1}\bfF(\l) \bfg^T(\l) d\l,   \eqn Chij.a. \cr
\hat{\chi}_j &= -\int_{\G} \l^{j-1}\bff(\l) \bfG^T(\l) d\l.  \eqn Chijhat.b.}
$$
These must of course also satisfy the algebraic recursion relations
$$
\hat{\chi}_j= - \chi_j - \sum_{k=1}^{j-1} \chi_{j-k} \hat{\chi}_{k}\, ,\,
j>1,\quad \hat{\chi}_1= -\chi_1.   \eq Chirec..
$$

In the next section, we specialize to certain particular choices of the
functions $\bff(\l), \ \bfg(\l)$ determining the Fredholm kernel, and support
curves $\G$, for which the computation of the Fredholm determinant can be
reduced to the solution of an associated isomonodromic deformation problem.
The resulting systems, together with the generalization discussed in 
Section 4, extend those obtained in \cite{IIKS}, \cite{P} via the
Riemann-Hilbert method and in \cite{JMMS}, \cite{HTW}, \cite{TW1--TW3} by
other related methods.

\medskip
\section 2.  Special Integrable Kernels and Isomonodromic Deformations
\medskip

In the following, we choose $n=2m$ to be even, and let 
$\{\a_j\}_{j=1\dots n}$ be a set of distinct points along $\G$, ordered
according to its orientation, and let $\{\G_j\}_{j=1, \dots m}$ be the
ordered sequence of connected  segments of $\G$ such that $\G_j$ has
endpoints $(\a_{2j-1}, \a_{2j})$. Let 
$$
B= \diag(\b_1, \cdots \b_r)  \eq..
$$
be a diagonal $r \times r$ matrix with distinct eigenvalues
$\{\b_a\}_{a=1\cdots r}$ and $\{\bff_j, \bfg_j\}_{j=1 \cdots m}$  a set of
fixed $r \times p$ rectangular matrices satisfying
$$
\bfg_j^T\bff_k=0, \quad \forall j, k.  \eq fgjnull..
$$
Define
$$
\eqalignno{
\bff_0(\l)&:= \sum_{j=1}^m \bff_j \th_j(\l), \eqn flzero.a.  \cr
\bfg_0(\l)&:= \sum_{j=1}^m \bfg_j \th_j(\l),   \eqn glzero.b.}
$$
\nextnumber
where  $\{\th_j(\l)\}_{j=1 \dots m}$ denote the characteristic functions of 
the curve segments  $\{\G_j\}$, and
$$
\eqalignno{
\bff(\l)&:=   \Psi_0(\l)\bff_0(\l),  \eqn fl.a.  \cr
\bfg(\l)&:= \left(\Psi_0^T(\l)\right)^{-1}\bfg_0(\l),  \eqn gl.b.}
$$
where
$$
\Psi_0(\l) := e^{\l B}  \eq vacpsi..
$$
is viewed as a ``vacuum'' wave function.

Let $\chi_{\pm}(\l)$ be the solution to the matrix Riemann--Hilbert problem
\MRHasympt, \MRHdisc corresponding to the choice \fl, \gl for the functions
$\bff(\l), \bfg(\l)$. Define the invertible $r\times r$ matrix valued 
function $$
\Psi(\l):= \chi(\l)\Psi_0(\l),  \eq Psidef..
$$
with limiting values $\Psi_{\pm}$ on either side of the segments of $\G$
$$
\Psi_{\pm}(\l):= \chi_{\pm}(\l)\Psi_0(\l).  \eq ..
$$
Then $\Psi_{\pm}$ satisfy the discontinuity conditions
$$
\Psi_-(\l)= \Psi_+(\l) H_0(\l)  \eq Psidisc..
$$
across these segments, where
$$
H_0(\l) := \bfI_r +2\pi i h_0(\l) = \exp 2\pi i h_0(\l),  \eq HHlzero..
$$
and
$$
h_0(\l):= \sum_{j=1}^n \bff_j \bfg_j^T\th_j(\l)  \eq hhlzero..
$$
has rank $\le p$, is piecewise contant along the segments $\G_j$ and, because
of \fgjnull, satisfies
$$
h_0(\l)h_0(\mu) =0, \quad \forall \ \l, \mu. \eq hzerolammu..
$$
In terms of these quantities, $H(\l)$ is given by
$$
H(\l) = \Psi_0(\l) H_0(\l) \Psi_0^{-1}(\l).  \eq HHl..
$$
From \Fdr, \Gdr and \Psidef, it follows that the matrix--valued
functions $\bfF(\l)$, $\bfG(\l)$  determining the resolvent kernel are given
by 
$$ \eqalignno{
\bfF(\l)&=\Psi(\l)\bff_0(\l), \eqn FPsi.a. \cr
\bfG(\l) &= \left(\Psi^T(\l)\right)^{-1} \bfg_0(\l),  \eqn GPsi.b.}
$$
where, again, the choice of $\Psi_{\pm}(\l)$ on the RHS is immaterial because 
of the orthogonality conditions \fgjnull.

\medskip
\Subtitle {2a. Isomonodromic families of operators: deformation of the
endpoints}

\nobreak
  In this section, we construct an isomonodromic family of operators of the
form \DDl, \NNl having the ``dressed'' wave function $\Psi(\l)$ defined in
\Psidef as kernel. This follows by examining the analytic properties of
$\Psi(\l)$ in a neighborhood of the singular points  
$\{\l=\a_j\}_{j=1 \dots n}$ and asymptotically as $\l\ra \infty$.
Specifically, define $\{N_j\}_{j=1\dots n}$  by
$$
N_j := -\bfF_j \bfG^T_j,  \eq Nj..
$$
where
$$
\bfF_j:= \lim_{\l \ra \a_j} \bfF(\l) \quad
\bfG_j:= (-1)^{j}\lim_{\l \ra \a_j} \bfG(\l),  \eq FGjdef..
$$
with the  limit $\l \ra \a_j$  taken inside the segment $\G_j$. We then have 
the following result.

\Proclaim Theorem 2.1: The wave function $\Psi(\l)$ defined by \Psidef
satisfies the equations
$$
\eqalignno{
{\di \Psi \over \di \l} -
\left(B  + \sum_{j=1}^n {N_j \over \l - \a_j}\right)\Psi &=0,
\eqn Psilambdaeq.a. \cr
{\di \Psi \over \di \a_j}+ {N_j \over \l -\a_j}\Psi&=0, \eqn Psialphaeq.b.}
$$
with the $N_j$'s given by \Nj. This implies the commutativity
$$
[\DD_\l, \ \DD_{\a_j}]=0,\quad   [\DD_{\a_i}, \ \DD_{\a_j}]=0,
\quad i, j=1, \dots n  \eq DDlalphacomm..
$$
of the operators
$$
\eqalignno{
\DD_\l &:= {\di \over \di \l} - B - \sum_{j=1}^n {N_j \over \l -\a_j}
={\di \over \di \l} - \NN(\l), \eqn Dlam.a. \cr
\DD_{\a_j} &:= {\di \over \di \a_j} + {N_j \over \l -\a_j}, \eqn Daj.b.}
$$
and hence the invariance of the monodromy data of the operator $\DD_\l$ under
changes in the parameters $\{\a_j\}$ .

\Demo Proof: The discontinuity conditions \Psidisc may be written
$$
\Psi_+(\l) - \Psi_-(\l) = -2\pi i \Psi_+(\l) \sum_{j=1}^m \bff_j \bfg^T_j
\th_j(\l).
\eq..
$$
From analyticity, and the fact that the $\bff_j \bfg_j^T$  satisfy
$$
\bff_j \bfg_j^T \bff_k \bfg_k^T =0, \quad \forall j,k,  \eq..
$$
which follows from \fgjnull, we conclude that, in a neighborhood of $\G_j$,
$\Psi(\l)$ can be written
$$
\Psi(\l) = \hat{\psi}_j(\l) \left( {\l - \a_{2j} \over
\l -\a_{2j-1}}\right)^{-\bff_j \bfg^T_j},
\eq Psijloc..
$$
where $\hat{\psi}_j(\l)$ is locally holomorphic and invertible. It follows 
that the logarithmic derivative $\Psi_\l \Psi^{-1}$ is meromorphic for finite
$\l$ with simple poles at the points $\{\l=\a_j\}_{j=1 \cdots n}$, and the
residue matrices are given by
$$
N_j := (-1)^{j+1}\hat{\psi}_{[{j+1\over 2}]}(\a_j) \bff_{[{j+1\over 2}]}
 \bfg_{[{j+1\over 2}]}^T \hat{\psi}_{[{j+1\over 2}]}^{-1} (\a_j).
\eq Njpsi..
$$
The local form $\Psijloc$ impies that these $N_j$'s are given by \Nj, with
$\bfF_j, \bfG_j$ defined by \FGjdef.
Note that, by eq. \FGnull, we also have
$$
\bfF^{T}_{j}\bfG_{j} = 0.
\eq FTGO..
$$

  From the asymptotic form \MRHasympt and the equation \Psidef
, we see that, for large $|\l|$, $\Psi_\l\Psi^{-1}$ is of the form
$$
\Psi_\l \Psi^{-1} \sim    B  + \OO(\l^{-1}),  \eq..
$$
and hence, by Liouville's theorem, $\Psi$ satisfies \Psilambdaeq.

  From the local form \Psijloc of $\Psi$ and the fact that $\Psi$ is analytic
away  from $\G$, we see that ${\di \Psi \over \di \a_j}\Psi^{-1}$ is
holomorphic away from  $\a_j$ and that in a neighborhood of $\a_j$, the sum
$$
{\di \Psi \over \di \a_j}\Psi^{-1} + {N_j \over \l -\a_j} \eq Psidif..
$$
is analytic. From the asymptotic form \MRHasympt and \Psidef it also follows
that ${\di \Psi \over \di \a_j}\Psi^{-1}$ vanishes at $\l=\infty$ and hence,
again by Liouville's theorem, the expression in \Psidif vanishes identically,
hence proving that $\Psi$ satisfies \Psialphaeq.

The fact that the invertible matrix $\Psi$ satisfies both \Psilambdaeq and
\Psialphaeq implies the commutativity conditions \DDlalphacomm
for the operators $\DD_\l$, $\DD_{\a_j}$  defined in \Dlam and \Daj.
These deformation equations for the operator  $\DD_\l$ imply the
invariance of its monodromy data both at  the finite regular singular points
$\{\a_j\}_{j=1\dots n}$ and at the irregular sigular point $\l=\infty$ (cf.
\cite{JMMS, JMU}).  \hfill Q.E.D.

   Equivalently, we may consider the equations for the matrices $\bfF_j$,
$\bfG_j$ implied by taking the limits $\l \ra \a_k$ in \Psilambdaeq,
\Psialphaeq.
\Proclaim Corollary 2.2: The matrices $\bfF_j$, $\bfG_j$ satisfy the
differential system
$$
\eqalignno{
{\di \bfF_j\over \di \a_k} &= -{N_k\over \a_j - \a_k}\bfF_j, \quad j \neq k,
\eqn Fjk.a. \cr
{\di \bfF_j\over \di \a_j} &=
\left(B +\sum_{k=1 \atop \k \neq j}^n {N_k \over \a_j - \a_k}\right)\bfF_j,
\eqn Fjj.b. \cr
{\di \bfG_j\over \di \a_k} &= {N^T_k\over \a_j - \a_k}\bfG_j, \quad j \neq k,
\eqn Gjk.c. \cr
{\di \bfG_j\over \di \a_j} &=
-\left(B +\sum_{k=1 \atop \k \neq j}^n {N^{T}_k \over \a_j -
\a_k}\right)\bfG_j.
\eqn Gjj.d. \cr}
$$

\Demo Proof: This follows from the definitions \FPsi, \GPsi, \FGjdef and
equations \Psilambdaeq, \Psialphaeq of the theorem.

\noindent {\it Remark.}
 Let $N=np$ and define the pair of rectangular $N\times r$ matrices $F,\ G$
from the $p\times r$ blocks $\{\bfF_j^T,\ \bfG_j^T\}$
$$
F:= \pmatrix{\bfF_1^T \cr \bfF_2^T \cr \cdot \cr \cdot \cr \bfF_n^T}
\quad
G:= \pmatrix{\bfG_1^T \cr \bfG_2^T \cr \cdot \cr \cdot \cr \bfG_n^T}.
\eq FGident..
$$
Then
$$
\NN(\l)= B + \sum_{j=1}^n {N_i \over \l -\a_j}=B+F^T(A-\l\bfI)^{-1}G,  \eq
NBFAG..
$$
as in \NNintr, where  $A$ is the diagonal $N\times N$ matrix with eigenvalues
$\{\a_j\}_{j=1 \cdots n}$ all of multiplicity $p$, and the equations 
\Fjk - \Gjj are just the Hamiltonian equations  \FHameqintr. 
\medskip

\Subtitle {2b. Deformations of exponents}

\nobreak
We now turn to the dependence of the wave function $\Psi(\l)$ on the 
parameters $\{\b_a\}_{a=1, \dots r}$. By examining the analytic structure of
the logarithmic derivatives ${\di \Psi \over \di \b_j}$ near $\G$ and
asymptotically, we can similarly derive differential equations for $\Psi$ 
with respect to these parameters, and hence deduce the independence of the 
monodromy of $\DD_\l$ of the parameter values.
\Proclaim  Theorem 2.3:  The wave function $\Psi(\l)$ satisfies the equations
$$
\DD_{\b_a} \Psi = 0, \quad a=1, \dots r, \eq DbetaPsi..
$$
where the operators $\{\DD_{\b_a}\}_{a=1, \dots r}$ are defined by
$$
\DD_{\b_a}:= {\di \over \di \b_{a}} - \l E_{a} -  \sum_{b=1 \atop b\neq a}^r
 {E_a \left(\sum_{j=1}^nN_j\right) E_b
+ E_b \left(\sum_{j=1}^nN_{j} \right)E_a \over \b_a - \b_b}, \quad a=1, 
\dots r,  \eq Dba..
$$
with the $N_i$'s given by \Nj, \FGjdef and $E_{a}$ given by \Eaintr. This
implies the commutativity
conditions
$$
\left[\DD_\l, \ \DD_{\b_a}\right] =0,  \quad
\left[\DD_{\b_a}, \ \DD_{\b_b}\right] =0,\quad a,b=1, \dots r,
\eq DlambdaDbeta..
$$
and hence invariance of the monodromy data of $\DD_\l$  under the
deformations  parameterized by $\{\b_a\}_{a=1, \dots r}$.

\Demo Proof:
Since the matrix $H_0(\l)$ entering the discontinuity conditions
\Psidisc is  independent of the parameters $\b_a$, we see that the 
logarithmic derivatives of $\Psi_{\pm}$ with  respect to these parameters
have no discontinuity across $\G$
$$ {\di \Psi_+ \over \di \b_a} \Psi_+^{-1} = {\di
\Psi_- \over \di \b_a} \Psi_-^{-1}. \eq..
$$
  From the local structure \Psijloc, it follows that
${\di \Psi \over \di \b_a}\Psi^{-1}$ is in fact holomorphic in a neighborhood 
of $\G$, and hence throughout the complex $\l$ plane. Furthermore, from the
asymptotic form  \chiasympt and \Psidef, it follows that the difference
$$
{\di \Psi \over \di \b_a}\Psi^{-1} - \left(\l E_a +
\left[\chi_1, \ E_a\right]\right)
\eq Psibasympt..
$$
vanishes in the limit $\l\ra \infty$. Therefore, again by Liouville's theorem
this difference, being globally holomorphic, vanishes identically
$$
{\di \Psi \over \di \b_a} - \left(\l E_a + \left[\chi_1, \ E_a\right]\right)
\Psi=0.
\eq Psibetaeq..
$$

  Since $\Psi$ is invertible, the compatibility of \Psibetaeq and  
\Psilambdaeq implies
$$
\left[{\di \over \di \l} - B - \sum_{j=1}^n {N_j \over \l -\a_j},\
{\di \over \di \b_a} - \l E_a -\left[\chi_1, \ E_a\right]\right]=0,
\quad a=1, \dots r.  \eq abcommut..
$$
From the asymptotic form of this equation, it follows that the off diagonal
terms of $\chi_1$ are given by
$$
(\b_b - \b_a)\left( \chi_1\right)_{ab} = \sum_{j=1}^n\left(N_j\right)_{ab},
\eq Psibetadef..
$$
and hence
$$
\left[\chi_1, E_a\right]_{bc} = (\d_{ab} -\d_{ac})
{\left(\sum_{j=1}^nN_j\right)_{bc} \over \b_b - \b_c}
\eq PsiEacom..
$$
Equation \Psibetaeq is therefore just the condition that $\Psi$ be in the 
joint kernel of the operators defined in \Dba and  \abcommut are the
commutativity conditions implying that the monodromy of $\DD_\l$ is 
preserved under the deformations  parameterized by $\{\b_a\}_{a=1, \dots r}$.
\hfill Q.E.D.

\noindent {\it Remark.} The compatibility of eq. \Psibetaeq with \Psialphaeq
also implies the commutativity conditions
$$
[\DD_{\a_i}, \ \DD_{\b_a}]=0, \quad i =1, \dots n, \ a =1, \dots r  \eq..
$$

  Finally, we may again equivalently consider the system of equations 
satisfied by the  quantities $\{\bfF_j, \ \bfG_j\}_{j=1\cdots n}$, which
follow from \Psilambdaeq and \DbetaPsi.

\Proclaim Corollary 2.4: The matrices $\bfF_j$, $\bfG_j$ satisfy the
differential system
$$
\eqalignno{
{\di \bfF_j \over \di \b_a}  & = \left(\a_j E_a + \left[ \chi_1, \
E_a\right]\right) \bfF_j, \eqn Fja.a. \cr
{\di \bfG_j \over \di \b_a}  & = \left( -\a_jE_a + \left[ \chi^T_1, \
E_a\right]\right) \bfG_j, \eqn Gja.b.}
$$
with the term $\left[ \chi_1, \ E_a\right]$ given by \PsiEacom,
which are exactly the Hamiltonian equations \GHameqintr under the
identifications \FGident.

\medskip
\Subtitle {2c. The Fredholm determinant}

\nobreak
We now proceed in a standard way to compute the logarithmic differential
\Freddetdiff with respect to the parameters 
$\{\a_j, \b_a\}_{1\leq j \leq n, \ 1\leq a \leq r}$. The result is 
summarized by the formula 
\Proclaim Theorem 2.5:
$$
d\ln \det (\bfI - \bfK)  = \o =\sum_{k=1}^n H_k d\a_k 
+ \sum_{a=1}^r K_ad\b_a,  \eq dettau..
$$
where the individual factors may be expressed
$$
\eqalignno{
H_k &= {\di \ln \det (\bfI - \bfK) \over \di \a_k}
= \tr(BN_k) + \sum_{j=1\atop j\neq k}^n{\tr(N_j N_k)\over \a_-\a_j},
 \eqn Hk.a. \cr
K_a &= {\di \ln \det (\bfI - \bfK) \over \di \b_a}=
 \sum_{j=1}^n\a_j\left(N_j\right)_{aa} + 
\sum_{b=1 \atop b\neq a}^r {(\sum_{j=1}^n N_j)_{ab}
(\sum_{k=1}^n N_k)_{ba} \over{\b_a-\b_b}}. \eqn Ka.b.}
$$

\Demo Proof:
To derive \Hk we use the equation
$$
\ln \det (\bfI - \bfK) = \tr \ln (\bfI - \bfK) = -\sum_{n=1}^{\infty}
{\tr \bfK^{n}\over n}. \eq detKexp..
$$
From the formula
$$
\tr \bfK^{n} = \tr \sum_{j_{1},...j_{n}=1}^{m}
\int_{\a_{2j_{1}-1}}^{\a_{2j_{1}}}
\dots \int_{\a_{2j_{n}-1}}^{\a_{2j_{n}}}K(\lambda_{1},\lambda_{2})
K(\lambda_{2},\lambda_{3})
\dots K(\lambda_{n},\lambda_{1})d\lambda_{1}d\lambda_{2}...d\lambda_{n}, 
\eq ..
$$
we conclude that
$$
{\di \tr \bfK^{n} \over \di \a_{k}} = (-1)^{k}n\tr K^{n}(\a_{k},\a_{k}),
\eq Kak..
$$
where $K^{n}(\l,\mu)$ stands for the kernel of $\bfK^{n}$ and ``$\tr $'' in
the r.h.s. is
the matrix trace.
Equations \detKexp and \Kak imply (cf. [TW]) that
$$
{\di \ln \det (1 - \bfK) \over \di \a_{k}} =  (-1)^{k+1}\tr
\sum_{n=1}^{\infty}K^{n} (\a_{k},\a_{k}) = (-1)^{k+1}\tr R(\a_{k},\a_{k}),
\eq..
$$
where the second equality follows from the Neumann expansion of the resolvent
operator $\bfR$.

From equations \Rdiag, \GPsi, and \Psilambdaeq it follows that
$$
R(\lambda,\lambda) = \bfF^{T}(\lambda)\left (
B + \sum_{j=1}^{n}{N_{j}^{T}\over {\lambda-\a_{j}}} \right )\bfG(\lambda) .
\eq RFG..
$$
Passing in the last equation to the limit $\lambda \rightarrow \a_{k}$ and
taking into account \FGjdef and \FTGO,
we conclude that
$$
R(\a_{k},\a_{k}) = (-1)^k\bfF^{T}_{k}\left (B + \sum_{j=1 \atop j\neq k}^{n}
{N_{j}^{T}\over {\a_{k}-\a_{j}}}\right )\bfG_{k},
\eq RFGk..
$$
which, because of \Nj, implies that
$$
\tr R(\a_{k},\a_{k}) = (-1)^{k+1}\left (\tr(BN_{k}) +
\sum_{j=1 \atop j\neq k}^{n}{\tr(N_{j}N_{k})\over {\a_{k}-\a_{j}}}\right),
\eq RNN..
$$
and hence \Hk.

To prove \Ka we  use \Freddetdiff, which implies
$$
K_{a} = - \tr \left ((1 - \bfK)^{-1}{\di \bfK \over \di \b_{a}}\right ).
\eq Ktwo..
$$
From \fl it follows that
$$
{\di K(\lambda,\mu) \over \di \b_{a}} = \bff^{T}(\lambda)E_{a}\bfg(\mu).
\eq diKba..
$$
Equations \Ktwo and \diKba imply that
$$
K_{a} = -\tr \int_{\G}\bfF^{T}(\mu)E_{a}\bfg(\mu)d\mu,  \eq ..
$$
which in view of \Chij  means that
$$
K_{a} = -\tr(E_{a}\chi_{1}) = - (\chi_{1})_{aa}.
\eq Kachi..
$$

From the asymptotic form of the equation \Psilambdaeq, it follows
(cf. \Psibetadef)  that
$$
\eqalignno{
[\chi_1,B] &=  \sum_{j=1}^n N_j, \eqn chiN.a. \cr
\chi_1 &= [\chi_{2},B] -\sum_{j=1}^n N_j\chi_1 -
\sum_{j=1}^n \a_j N_j. \eqn chiNN.b.}
$$
The last two formulae determine the diagonal elements of $\chi_{1}$ as
$$
(\chi_{1})_{aa} = - \sum_{j=1}^{n}\a_{j}(N_{j})_{aa}-
\sum_{b=1 \atop b\neq a}^r
{{(\sum_{j=1}^{n}N_{j})_{ab}(\sum_{k=1}^{n}N_{k})_{ba}}
\over{\b_{a}-\b_{b}}}, \eq chiaa..
$$
from which follows the expression \Ka for $K_a$.
\hfill Q.E.D.

\medskip
\Subtitle {2d. Monodromy Data and the Fredholm Determinant as a 
$\t$--Function}

\nobreak
Let
$$
{\di\Psi(\l)\over \di\l} =
\left ( B + \sum_{j=1}^{n}{A_{j}\over{\l-\a_{j}}}\right) \Psi(\l) \eq PsiA..
$$
be an arbitrary $r\times r$ linear system having the same singularity
structure  as the system \Psilambdaeq. According to [JMU], the $\t$-function
associated to a  given solution
$$
A_{j} = A_{j}(\a_{1},...,\a_{n};\b_{1},...,\b_{r})
\eq ..
$$
of the corresponding isomonodromic deformation
equations is defined via the differential form (see formula (5.17) in
[JMU]): 
 $$
\eqalign{
\o = & {1\over 2}\sum_{j\neq k}\tr{A_{j}A_{k}}{{d\a_{j}-d\a_{k}}
\over{\a_{j}-\a_{k}}} +
\sum_{a=1}^{r}\sum_{j=1}^{n}(A_{j})_{aa}d(\a_{j}\b_{a}) \cr
&+{1\over 2}\sum_{a\neq b} (\sum_{j=1}^{n}A_{j})_{ab}
(\sum_{j=1}^{n}A_{j})_{ba} {{d\b_{a}-d\b_{b}}\over{\b_{a}-\b_{b}}}.}
\eq oA..
$$
More precisely,  it is shown in [JMU]  (and also follows from the Hamiltonian
structure of the deformation equations [H]) that the form \oA is closed,
allowing one to introduce the $\t$--function, defined up to multiplication 
by local contants by the equation:
$$
d\ln \t = \o.
\eq logtau..
$$

The monodromy data corresponding to system \PsiA consist of the set of $n+1$
monodromy matrices,
$$
\bfM_{j},\quad j=1,..., n,\quad \bfM_{\infty},
\eq ..
$$
$$
\bfM_{\infty}\bfM_{n}\bfM_{n-1}...\bfM_{1}=\bfI_{r},
\eq ..
$$
associated to the singular points,
$$
\a_{j},\quad j=1,..., n.\quad \a_{\infty} = \infty,
\eq ..
$$
augmented by the two Stokes matrices,
$$
\bfS_{1},\, \bfS_{2}
\eq ..
$$
related to $\a_{\infty}$, which is the only irregular singular
point of \PsiA. The Riemann-Hilbert problem \Psidisc provides a special
solution of the isomonodromy equations,
$$
A_{j}(\a_{1},...,\a_{n};\b_{1},...,\b_{r}): = N_{j},
\eq AN..
$$
where $N_{j}$ is defined in \Nj. This solution is characterized by the
following  choice of the monodromy data (cf. \Psijloc):
$$
\eqalignno{
\bfM_{2j} &= \bfM^{-1}_{2j-1} = \bfI_{r} - 2\pi i \bff_{j}\bfg^{T}_{j} =
\exp(- 2\pi i \bff_{j}\bfg^{T}_{j}),\quad j = 1,...,m,\quad n=2m,
\eqn Mspec.a. \cr
\bfM_{\infty} &= \bfS_{1}=\bfS_{2}= \bfI_{r}.
\eqn MSspec.b.}
$$

Comparing now the equations \dettau and \oA (cf. also eqs.
\omegintr--\tauintr), we conclude that the  $\t$-function evaluated on the
solution \AN of the isomonodromy equations is given by  the Fredholm
determinant, $$
 \t(\a_1, \dots, a_n;\b_1 \dots \b_r) = \det(\bfI - \bfK)
\eq Kdettau..
$$
of the integrable Fredholm operator $\bfK$  characterized by the
choice \fl, \gl for the functions $\bff(\l), \bfg(\l)$ in \Kdef. It is also
worth noticing,
that the equation \Kdettau follows directly from the equations \omegintr
-\tauintr.

 We now show that this Fredholm determinant is also a $\t$--function in the
sense of Segal--Wilson \cite{SW} and Sato \cite{Sa}. For this, we assume $\G$
is a simple, closed curve containing the origin of the $\l$--plane in its
interior. (In [SW], $\G$ is taken as the unit circle, but this is of no
significance to the general formulation.) Following \cite{SW}, for any 
positive integer $k$, let $\HH^k = L^2(\G, \bfC^k)$ be the Hilbert space of
square integrable $\bfC^k$--valued functions on $\G$,  and let
$$
\HH^k = \HH^k_+ + \HH^k_-  \eq Hpm..
$$
be its decomposition as the direct sum of subspaces $\HH_+^k$, $\HH_-^k$
consisting, respectively, of elements admitting a holomorphic extension to
the interior ($+$) or the exterior ($-$) of $\G$, the latter normalized to
vanish  as $\l\ra\infty$. Now, take $k=r$, and let $W\ss \HH^r$ be a subspace
that is the graph of a compact linear operator
$$
h_W:\HH_+^r \ra \HH_-^r.  \eq..
$$
Let
$$
\g(\bft): \HH^r \lra \HH^r, \quad \g \in \G_+^r  \eq..
$$
denote the action of an abelian subgroup $\G^r\ss \LGl(r)_+$ of the loop
group $\LGl(r)_+$ of nonsingular $r \times r$ matrix valued functions on $\G$
admitting a holomorphic nonsingular extension to the interior of $\G$, acting
by left multiplication on  $\HH^r$. Let $W(\bft)$ be the image of $W$
under the action of $\g(\bft)$, and choose (in the sense of \cite{SW}) an
admissible basis, in which $W$ is represented by the $(2\infty) \times
\infty$ matrix 
$$
W \sim \pmatrix {w_+ \cr w_-},  \eq..
$$
where $w_+, \ w_-$ represent maps
$$
w_+:\HH^r_+ \ra \HH^r_+, \qquad w_-:\HH^r_+ \ra \HH^r_-  \eq..
$$
the summed images of which give $W$. In this notation, we have
$$
w_- = h_W \circ w_+  \eq..
$$
The action of $\g(\bft)$ can then be expressed in matricial form as
$$
W(\bft)=\g(\bft) W = \pmatrix {a(\bft) & b(\bft) \cr 0 & d(\bft)}
\pmatrix{w_+ \cr w_-} = \pmatrix {a(\bft) w_+ + b(\bft)w_- \cr d(\bft) w_-}.
\eq Wgammat..
$$
where the maps
$$
\eqalignno{
a(\bft):\HH_+^r &\ra \HH_+^r  \eqn.a.\cr
b(\bft):\HH_-^r &\ra \HH_+^r  \eqn.b.\cr
d(\bft):\HH_-^r &\ra \HH_-^r  \eqn.c.}
$$
are defined by restriction of $\g(\bft)$ to $\HH_{\pm}^r$ composed with the
orthogonal projection maps
$$
P_{\pm}:\HH^r \ra \HH_{\pm} \eq..
$$
$$
a:=P_+ \circ \g \vert_{\HH_+^r}, \quad
b:=P_+ \circ \g \vert_{\HH_-^r}, \quad
d:=P_- \circ \g \vert_{\HH_-^r}.
$$
A convenient way to express these projection maps is via the Cauchy operators
$$
(P_{\pm} v)(\l) = \pm {1\over 2\pi i}\lim_{\l_{\pm} \ra \l}
\oint_\G {v(\mu)\over \mu -\l_{\pm}}d\mu
=:\pm {1\over 2\pi i}\oint_\G {v(\mu)\over \mu -\l_{\pm}}d\mu,  \eq Cauchy..
$$
where the expression on the right is an abbreviated notation for the limit,
indicating that after integration $\l_{\pm}$ approaches  $\l\in \G$ from the
corresponding side $\G_{\pm}$. The Segal--Wilson--Sato $\t$--function for 
such a subspace $W\ss \HH^r$ is then just given by the determinant
 $$
\t_W(\bft) := \det\left(\bfI + b(\bft) \circ h_W \circ a(\bft)^{-1}\right).
  \eq taufunc..
$$

   For the specific case at hand, we let $\G$ be a closed
curve passing through the points $\{\a_i\}_{i=1\dots n}$ and define $W$ to
be the image of $\HH_+^r$ under the action of the loop group element
$H^{-1}_0(\l)$ where
$H_0(\l)$ is defined in  \HHlzero
 $$
W=W(\a_1, \dots \a_n) := H^{-1}_0(\HH_+^r) = \{ v_+ - 
2\pi i h_0 v_+, \ v_+\in \HH_+^r\}.  \eq..
$$
The abelian subgroup $\G_+^r \ss \LGl(r)$ is taken to be the set of values 
of the ``vacuum'' wave function
$$
\g(\bft) := \Psi_0(\b_1, \dots \b_r) = e^{\l B},
\quad \{t_a:=\b_a\}_{a=1, \dots r}.   \eq..
$$
Then, using \Cauchy, and the orthogonality condition \hzerolammu, we conclude
that $$
W=\left\{ v_+ +\int_\G {h_0(\mu) v_+(\mu)\over \mu -\l_-} d\mu, \ v_+\in
\HH_+^r\right\}.  \eq..
$$
Hence  we may express the
operator $h_W$ as
$$
h_W v_+(\l) = \int_\G {h_0(\mu) v_+(\mu)\over \mu -\l_-} d\mu,   \eq hWmap..
$$
while the maps $a$, $b$ and $d$ are given by
$$
\eqalignno{
a v_+(\l) &= \Psi_0 (\l) v_+(\l) \eqn amap.a.  \cr
b v_-(\l) &= {1\over 2 \pi i} \int_\G {\Psi_0(\mu) v_-(\mu) \over
\mu - \l_+} d\mu \eqn bmap.b.\cr
d v_-(\l) &= -{1\over 2 \pi i} \int_\G {\Psi_0(\mu) v_-(\mu) \over
\mu - \l_-} d\mu \eqn dmap.c.\cr
v_+ \in \HH_+^r,& \quad v_-\in \HH_-^r.}
$$

We then have the following theorem.
\Proclaim Theorem 2.6: The following two determinants are equal
$$
\det(\bfI - \bfK) =
\det\left(\bfI + b(\bft) \circ h_W \circ a(\bft)^{-1}\right) ,  \eq
tautheorem..
$$
and hence the $\tau$--function $\tau(\a_1, \dots \a_n;, \b_1, \dots \b_r)$
evaluated in \Kdettau
is the same as the Sato--Segal--Wilson $\tau$--function $\t_W(\bft)$ defined 
in \taufunc.

\Demo Proof:  Composing the inverse $a^{-1}$ of the map defined in \amap 
with $h_W$ and $b$ as defined in \hWmap, \bmap gives the map
$b\circ h_W \circ a^{-1} : \HH_+^r \ra \HH_+^r$ as
$$
b\circ h_W \circ a^{-1} ( v_+ )(\l) =
 {1\over 2\pi i} \oint_\G \oint_\G {\Psi_0(\nu) \over \nu - \l_+}{ h_0(\mu)
\Psi_0^{-1}(\mu) v_+(\mu) \over \mu - \nu_-} d\mu d\nu.  \eq detKfg..
$$
Inverting the order of integration, evaluating the residues at $\nu = \mu$
and $\nu =\l_+$ and using \hhlzero gives
$$
b\circ h_W \circ a^{-1} ( v_+ )(\l) =
-\oint_\G {\bff(\mu) \bfg^T(\mu) v_+(\mu) \over \mu - \l_+}d\mu +
\oint_\G {\Psi_0(\l)\bff_0(\mu) \bfg^T(\mu) v_+(\mu) \over \mu -
\l_+}d\mu,  \eq bhWainv..
$$
with $\bff(\mu)$, $\bfg(\mu)$ defined by \fl, \gl.
Now note that $\bfK:\HH^p \ra \HH^p$ can be written in the
form
$$
\bfK = \bfA_{\bff} \circ \bfB_{\bfg},  \eq KAfBg..
$$
where $\bfB_{\bfg}: \HH^p\ra \HH_+^r$ is defined by
$$
\bfB_{\bfg} \phi(\l) = \oint_{\G} {\bfg(\mu) \phi(\mu)\over \l-\mu}d\mu,
\quad \phi \in \HH^p,   \eq Bsubg..
$$
while $\bfA_{\bff}:\HH_+^r \ra \HH^p$ is defined by
$$
\bfA_{\bff}  v_+(\l) = \bff^T(\l)v_+(\l), \quad v_+ \in \HH_+^r .
\eq Asubf..
$$
On the other hand, from \bhWainv, we see that $b\circ h_W \circ a^{-1}$
can be expressed as
$$
b\circ h_W \circ a^{-1} = \wt{\bfB_{\bff}}\circ \bfA_{\bfg}  \eq..
$$
where
$$
\wt{\bfB_{\bff}} = \bfB_{\bff} - a\circ\bfB_{\bff_0} \eq..
$$
with $\bfB_{\bff}$ and $\bfB_{\bff_0}$ defined as in \Bsubg, with
$\bfg$ replaced by $\bff$ and $\bff_0$, repectively (cf. \fl, \flzero),
while $\bfA_{\bfg}$ is defined as in \Asubf, with $\bff$ replaced by
$\bfg$.

The Weinstein--Aronszajn identity implies that
$$
\det(\bfI + \wt{\bfB_{\bff}}\circ \bfA_{\bfg}) =
\det (\bfI  + \bfA_{\bfg}\circ \wt{\bfB_{\bff}})
= \det (\bfI  +\bfA_{\bfg} \circ \bfB_{\bff}),
  \eq WeAr..
$$
where the second equality follows from that fact that, due to the 
orthogonality conditions \fgjnull, we have
$$
\bfA_{\bfg} \circ a \circ \bfB_{\bff_0} =  \bfA_{\bfg_0}\circ \bfB_{\bff_0}
=0.
  \eq..
$$
Therefore
$$
\t_W(\bft) =\det(\bfI + b(\bft) \circ h_W \circ a(\bft)^{-1})
=\det (\bfI - \bfK^T),  \eq detKgf..
$$
where
$$
\bfK^T := -\bfA_{\bfg} \circ \bfB_{\bff}.  \eq KAgBf..
$$
Comparing \KAfBg with \KAgBf, we see that $\bfK^T$ is really just
the transpose of $\bfK$, and hence the determinants in \tautheorem and 
\detKgf are the same.
\hfill Q.E.D.

\noindent
{\it Remark}. An identity related to \WeAr; namely, the equation
$$
(\bfI + \wt{\bfB_{\bff}}\circ \bfA_{\bfg})^{-1} +
\wt{\bfB_{\bff}}(\bfI  + \bfA_{\bfg}\circ \wt{\bfB_{\bff}})^{-1}\bfA_{\bfg}
= \bfI,
$$
lies, in fact, behind the basic relationship  \Fdr ,  \Gdr between the
Riemann-Hilbert problem and integrable Fredholm kernels (see \cite{DIZ}, 
proof of Lemma 2.12).
\section 3. Duality
\medskip \nobreak

\Subtitle {3a. The Dual Isomonodromic System}

\nobreak
  In the following, choose $r$ to be even, $r=2s$, and let
$\wt{\G}$ be a piecewise continuous oriented curve in the complex $z$--plane
consisting of the ordered sequence $\{ \wt{\G}_a\}_{a=1,\dots s}$ of 
connected components whose endpoints are given by the consecutive pairs
$\{\b_{2a-1}, \b_{2a}\}_{a=1.\dots s}$ of eigenvalues of the $qr\times qr$
matrix $B=\diag(\b_1, \dots
\b_{2s})$, where each eigenvalue $\b_a$ has multiplicity $q$, with $q\le n$.
Choose a set $\{\tbf_a,\tbg_a\}_{a=1,\dots s}$ of fixed rectangular 
$n\times q$ matrices satisfying
$$
\tbf_a^T\tbg_b =0, \quad  a,b=1, \dots , s  \eq..
$$
Let
$$
\eqalignno{
\tbf(z)&=e^{zA}\sum_{a=1}^s \tbf_a \tth_a(z), \eqn ftz.a. \cr
\tbg(z) &= e^{-zA} \sum_{a=1}^s \tbg_a \tth(z),   \eqn gtz.b.}
$$
where $\tth_a$ is now the characteristic function, along $\wt{\G}$, of the
curve segment $\wt{\G}_a$ and $A$ is the diagonal $n\times n$ matrix
$$
A= \diag(\a_1, \cdots \a_n).  \eq..
$$
Let $\tchi(z)$ be the nonsingular $n\times n$ matrix
valued function satisfying the conditions that it be analytic in the
compactified complex $z$--plane in the complement of $\wt{\G}$, with
asymptotic form
$$
\tchi(z) =\bfI_n + O(z^{-1}),  \eq..
$$
having cut discontinuities across $\wt{\G}$ given by
$$
\tchi_-(z) = \tchi_+(z)\tH(z), \quad z \in \wt{\G},   \eq..
$$
where $\tchi_+(z)$, $\tchi_-(z)$ are the limits as $\wt{\G}$ is approached 
from the left and right, respectively, and $\tH(z)$ is the invertible
$n\times n$ matrix function defined along $\wt{\G}$ by
$$
\tH(z) =\bfI_n + 2 \pi i \tbf(z) \tbg^T(z) = \exp 2\pi i \tbf(z) \tbg^T(z).
\eq..
$$
As above, we define the $q\times q$ matrix Fredholm integral operator $\tbK$
acting on $\bfC^q$--valued functions $\tbv$ on $\wt{\G}$ by
$$
\tbK(\tbv)(z) = \int_{\wt \G} \tK(z,w)\tbv(w)dw,  \eq..
$$
where
$$
\tK(z,w) = {\tbf^T(z) \tbg(w) \over z-w}.  \eq..
$$
The resolvent operator
$$
\tbR = (\bfI - \tbK)^{-1}\tbK  \eq..
$$
may again be expressed as
$$
\tbR(\tbv)(z) = \int_{\wt{\G}} \tR(z,w)\tbv(w), \eq..
$$
where the kernel
$$
\tR(z,w) ={\tbF^T(z)\tbG(w)\over z-w}  \eq..
$$
is determined by
$$
\eqalignno{
\tbF^T &= (\bfI-\tbK)^{-1}\tbf^T
 = (\bfI + \tbR)\tbf^T =\tbf^T \tchi^T, \eqn tFfK.a. \cr
 \tbG &= \tbg (\bfI-\tbK)^{-1}
=\tbg(\bfI + \tbR) = (\tchi^{T})^{-1}\tbg. \eqn tGgK.b.}
$$
\nextnumber
As above, $\tchi(z), \tchi^{-1}(z)$ have the asymptotic expansions
$$
\eqalignno{
\tchi(z) &= \bfI_n + \sum_{j=1}^{\infty} {\tchi_j\over z^j},  \eqn.a. \cr
\tchi^{-1}(z) &= \bfI_n + \sum_{j=1}^{\infty} {\hat{\tchi}_j\over z^j},
 \eqn.b.}
$$
\nextnumber
where
$$
\eqalignno{
\tchi_j &= \int_{\wt{\G}} z^{j-1}\tbF(z) \tbg^T(z) dz,   \eqn tChij.a. \cr
\hat{\tchi}_j &= -\int_{\wt{\G}} z^{j-1}\tbf(z) \tbG^T(\l) dz.
\eqn tChijhat.b.}
$$
satisfy
$$
\hat{\tchi}_j= - \tchi_j - \sum_{k=1}^{j-1} \tchi_{j-k} \hat{\tchi}_{k}\, ,
\quad j>1,\quad \hat{\tchi}_1= -\tchi_1.   \eq tChirec..
$$

  Again, we may define the $n \times q$ rectangular matrices
$$
\tbF_a:= \lim_{z \ra \b_a} \tbF(z) \quad
\tbG_a:= (-1)^{a}\lim_{z \ra \b_a} \tbG(z), \quad a=1, \dots r,  
\eq tFGjdef..
$$
where the limits are taken inside the curve segments
$\{\wt{\G}_a\}_{a=1, \dots s}$, the $n \times n$ matrices
$$
M_a := -\tbF_a \tbG_a^T, \quad a=1, \dots r.  \eq..
$$
and the pair of $n \times rq$  matrices $\wt{F}$, $\wt{G}$  formed from the
blocks  $\{\tbF_a,\ \tbG_a\}$
$$
\eqalignno{
\wt{F}&:= \pmatrix{\tbF_1 & \tbF_2 & \cdot & \cdot & \tbF_r},
\eqn tFident.a. \cr
\wt{G}&:= \pmatrix{\tbG_1 & \tbG_2 & \cdot & \cdot & \tbG_r}.
\eqn tGident.b.}
$$
Then
$$
\MM(z):= A + \sum_{a=1}^r {M_a \over z -
\b_a}=A+\wt{F}(B-z\bfI)^{-1}\wt{G}^T  
\eq MAFBG..
$$
as in \DDzintr, \MMintr.

Now define, similarly to the previous section,
$$
\tPsi(z):= \tchi(z)e^{z A},  \eq tPsidef..
$$
with limiting values on either side of the segments of $\wt{\G}$
$$
\tPsi_{\pm}(z):= \tchi_{\pm}(z)e^{z A}.  \eq ..
$$
It then follows from the same arguments as in the preceding section that 
$\tPsi$ simultaneously satisfies
$$
\eqalignno{
\wt{\DD}_z \tPsi & =0, \eqn.a. \cr
\wt{\DD}_{\b_a} \tPsi & =0, \quad a=1, \dots r,  \eqn.b. \cr
\wt{\DD}_{\a_i} \tPsi & =0, \quad i=1, \dots n, \eqn.c. \cr}
$$
\nextnumber
where
$$
\eqalignno{
\wt{\DD}_z &:= {\di \over \di z} - \MM(z) =
    {\di \over \di z} - A - \sum_{a=1}^r {M_a\over z-\b_a}, \eqn Dzt.a. \cr
\wt{\DD}_{\b_a} &:={\di \over \di \b_a} + {M_a\over z-\b_a}, \eqn Dat.b. \cr
\wt{\DD}_{\a_i}&:= {\di \over \di \a_{i}} - z E_{i} -
\sum_{j=1 \atop j\neq i}^n {E_i \left(\sum_{a=1}^r M_a\right) E_j
+ E_j \left(\sum_{a=1}^r M_{a}\right) E_i \over \a_i - \a_j},
  \eqn Dit.c.}
$$
\nextnumber 
with $\{E_i\}_{i=1,\dots n}$ the elementary $n\times n$ matrices with entry 
$1$ in the $(ii)$ position. Consequently, all these operators commute  and 
the monodromy of the parametric family of operators $\wt{\DD}_z$ is invariant
under the changes in the deformation parameters $\{\a_i, \b_a\}_{i=1\dots n,
\ a=1,  \dots r}$. The corresponding dynamical equations, expressed in terms
of the  matrices $\{\tbF_a, \tbG_a\}_{a=1, \dots r}$ are 
$$
\eqalignno{
{\di \tbF_a\over \di \b_b} &= -{M_b\over \b_a - \b_b}\tbF_a, \quad a \neq b,
\eqn tFjk.a. \cr
{\di \tbF_a\over \di \b_a} &=
\left(A +\sum_{b=1 \atop b \neq a}^r {M_b \over \b_a - \b_b}\right)\tbF_a,
\eqn tFjj.b. \cr
{\di \tbG_a\over \di \b_b} &= {M^T_b\over \b_a - \b_b}\tbG_a, \quad a \neq b,
\eqn tGjk.c. \cr
{\di \tbG_a\over \di \b_a} &= -\left(A +\sum_{b=1 \atop b \neq a}^r {M^{T}_b
\over \b_a - \b_b}\right)\tbG_a, \eqn tGjj.d. \cr
{\di \tbF_a \over \di \a_i}  & = \left( \b_a E_i + \left[ \tchi_1,\
E_i\right]\right)
\tbF_a, \eqn tFai.e. \cr
{\di \tbG_a \over \di \a_i}  & = \left( -\b_a E_i + \left[ \tchi^T_1, \
E_i\right]\right) \bfG_a, \eqn Gai.f.}
$$
with the term $\left[ \chi_1, \ E_i\right]$ given by
$$
\left[\tchi_1, E_i\right]_{jk} = (\d_{ij} -\d_{ik})
{\left(\sum_{a=1}^rM_a\right)_{jk} \over \a_j - \a_k}.
\eq PsiEicom..
$$
These are again just the Hamiltonian equations \FHameqintr, \GHameqintr under
the identifications \tFident, \tGident, provided the pair $(\tbF, \tbG)$ is
identified with $(\bfF, \bfG)$.

Finally, we may compute the logarithmic differential of the Fredholm
determinant
of $\tbK$ to obtain, as in Section 2c,
$$
d\ln \det (1 - \tbK)  = \tilde{\o} =\sum_{j=1}^n \tH_j d\a_j + \sum_{a=1}^r
\tK_a d\b_a,   \eq tdettau..
$$
where
$$
\eqalignno{
\tK_a &= {\di \ln \det (\bfI - \tbK) \over \di \b_{a}}
= \tr(A M_{a}) + \sum_{b=1 \atop b\neq a}^{r}{\tr{M_{a}M_{b}}\over
 {\b_{a}-\b_{b}}} \eqn tHk.a. \cr
\tH_j &= {\di \ln \det (\bfI - \tbK) \over \di \a_{j}}
=  \sum_{a=1}^{r}\b_{a}\left(M_{a}\right)_{jj} + \sum_{k=1 \atop k\neq j}^n
{ {\left (\sum_{a=1}^{r}M_{a}\right )_{jk}\left (\sum_{b=1}^{r}M_{b}\right
)_{kj}} \over{\a_{j}-\a_{k}}}.
\eqn tKa.b.}
$$
\medskip
\goodbreak
\Subtitle {3b. Duality Theorem}

\nobreak
   In order to relate the above isomonodromic deformation system to the one 
in Section 2 we henceforth make the following restrictions. First of all, we
consider  only the case $p=q=1$, so $\bfK$, $\tbK$ are both scalar Fredholm
integral operators. The rectangular matrices 
$\bff(\l), \bfg(\l), \bff_i, \bfg_i$ all become $r$-component column vectors
while $\tbf(z), \tbg(z), \tbf_a, \tbg_a$ are $n$-component column vectors.
Furthermore, the $\bff_i$'s and $\tbf_a$'s are chosen  to be of the special
form 
$$
\bff_j =  \bfe, \quad \tbf_a =  \tbe,
\quad j=1, \dots m, \quad a=1, \dots s,  \eq fjspec..
$$
where $\bfe \in \bfC^r$, $\tbe \in \bfC^n$ are the column vectors with all
components equal to $1$, while the components of the column vectors $\bfg_j$
and $\tbg_a$ are chosen such that
$$
(\bfg_{j})_{2a} = -(\bfg_{j})_{2a-1}
=  (\tbg_{a})_{2j} = -(\tbg_{a})_{2j-1} =: c_{ja}  \eq gjspec..
$$
for some constant $m\times s$ matrix with elements
$\{c_{ja}\}_{j=1,\dots m, a=1 \dots s}$.

   We then have the following result, which shows that the isomonodromic
deformation system \Dzt-\Dit is, in fact, the dual system to the one defined 
in Section 1 by eqs. \Dlam, \Daj, \Dba.

\Proclaim Theorem 3.1:
$\tbK$ is the Fourier-Laplace transform of $\bfK$ along the curves $\G$ and
$\wt{\G}$ and
$$
\eqalignno{
F&=\wt{F}, \quad G= \wt{G},\eqn dualFG.a. \cr
 \o &= \wt{\o}  \eqn dualomega.b.}
$$

\Demo Proof:
On the product $\G \times \wt{\G}$, define the following locally constant
function
$$
\hat{K}(\l, z) := \sum_{j=1}^m \sum_{a=1}^s c_{ja}\th_j(\l) \tth_a(z). \eq..
$$
Taking the Fourier--Laplace transform with respect to the variables $z$
and $\l$ along the curves $\wt{\G}$ and $\G$, respectively, gives the two
Fredholm kernels
$$
\eqalignno{
K(\l, \mu)& = \int_{\wt{\G}}\hat{K}(\mu, z) e^{z(\l -\mu)}dz
= {\bff^T(\l)\bfg(\mu)\over \l - \mu},   \eqn KFT.a. \cr
\wt{K}(w, z) &= \int_{\G}\hat{K}(\mu, z) e^{\mu(w-z)}d\mu
= {\tbf^T(w) \tbg(z) \over w-z}, \eqn tKFT.b.}
$$
where $\bff$, $\bfg$, $\tbf$, $\tbg$ are defined by \fl, \gl \ftz, \gtz
with $\bff_j$, $\bfg_j$ $\tbf_a$, $\tbg_a$ given by \fjspec, \gjspec.

To prove \dualFG, we use the Neumann expansion of the resolvent
$\bfR$ in \FfK to get the following convergent series for the
components of $F$
$$
\eqalignno{
\bfF_a(\a_j) =& \bff_a(\a_j) + \int_{\G}K(\a_j, \l_{1}) \bff_a(\l_1) d\l_1
+ \int_{\G} \int_{\G} K(\a_j,\l_2) K(\l_2, \l_1) \bff_a(\l_1) d\l_1 d\l_2 +
\cdots \cr
=&e^{\a_j\b_a} + \int_{\G} \int_{\wt{\G}} \hat{K}(\l_1, z_1) 
e^{z_1(\a_j -\l_1)} e^{\l_1 \b_a} dz_1 d\l_1 \cr
&+
\int_{\G}\int_{\G}\int_{\wt{\G}}\int_{\wt{\G}} \hat{K}(\l_2, z_2)
\hat{K}(\l_1, z_1)
 e^{z_2(\a_j -\l_2)} e^{z_1(\l_2 -\l_1)} e^{\b_a \l_1}dz_1 dz_2 d\l_1 d\l_2
+ \cdots  \cr
=&e^{\a_j\b_a} + \int_{\wt{\G}}\int_{\G}  \hat{K}(\l_1, z_1)
e^{\l_1(\b_a -z_1)} e^{z_1 \a_j} d\l_1 dz_1 \cr
&+
\int_{\wt{\G}}\int_{\wt{\G}}\int_{\G}\int_{\G} \hat{K}(\l_2, z_2) \hat{K}
(\l_1, z_1) e^{\l_2(z_1 -z_2)} e^{\l_1(\b_a -z_1)} e^{\a_j z_2}
d\l_1 d\l_2dz_1 dz_2  + \cdots  \cr
=&\tbf_j(\b_a) + \int_{\wt{\G}}{\wt{K}}(\b_a, z_1) \tbf_j(z_1) dz_1
+ \int_{\wt{\G}}\int_{\wt{\G}} {\wt{K}}(\b_a,z_1) {\wt{K}}(z_1, z_2)
\tbf_j(z_2) dz_1 dz_2 +
\cdots \cr
=&\tbF_j(\b_a),  \eqn..}
$$
where the Neumann expansion of $\tbR$ in \tFfK is used in the last line.
Similarly, using the Neumann expansions of $\bfR$ and $\tbR$ in \GgK, \tGgK 
and taking into account that
$$
\left( \sum_{k=1}^{m}c_{k[{a+1 \over
2}]}\theta_{k}(\l)\right)\hat{K}(\a_{j}, z) = \hat{K}(\l, \b_a)
\left( \sum_{b=1}^{s}c_{[{j+1 \over 2}]b}\theta_{k}(z)\right)
\eq ..
$$
$$
a=1,\dots r,\qquad j=1,\dots n,
$$
gives the equality
$$
\bfG_a(\a_j) = \tbG_j(\b_a). \eq..
$$
The equality \dualomega follows directly from the  fact that $\bfK$ and 
$\tbK$ are related by the Fourier-Laplace transforms \KFT, \tKFT. It may
also be seen from the equalities \dualFG in view of \NBFAG , \MAFBG and the
general duality relations mentioned in the introduction and proved in
\cite{H}. \hfill Q.E.D.

\section 4. Reductions and Generalizations  \medskip  \nobreak

\Subtitle {4a. Symplectic reduction}

\nobreak
  In applications, the structure of the Fredholm integral operator $\bfK$
and the associated Riemann--Hilbert problem data may be such that the generic
systems appearing in sections 2a, 2b are reduced to systems having fewer
independent variables. An important example of this involves reduction from 
the group $\grGl(2s)$ to the symplectic subgroup $\grSp(2s)$. Let $J$ be the
symplectic $2s \times 2s$ matrix having block form
$$
J= \pmatrix{ {\bf 0} & \bfI_s \cr -\bfI_s  & {\bf 0}},  \eq Jmat..
$$
and suppose that the matrix $B$ of section 2 satisfies the relations
$$
B^T J + J B =0,    \eq Bsympl..
$$
implying it is in the symplectic subalgebra $\grsp(2s) \ss \grgl(2s)$.
Since $B$
is diagonal, with eigenvalues $\{\b_a\}_{a=1,\dots 2s}$, this just means 
that 
$$
\b_{a+s} = - \b_a, \quad a=1, \dots s,  \eq..
$$
so $B$ has the form
$$
B = \pmatrix{ B_1 & {\bf 0} \cr {\bf 0} & -B_1}  \eq Bred..
$$
where $B_1= \diag(\b_1, \dots \b_s)$.
This implies that the vacuum wave function $\Psi_0(\l)$ satisfies
$$
\Psi_0^T(\l) J \Psi_0(\l) = J,  \eq Psizerosympl..
$$
and hence takes values in $\grSp(2s)$.
We also require that the $r \times p$ matrices $\{\bff_j, \bfg_j\}$ entering 
in \fl, \gl  satisfy the relations
$$
\bfg_j \bff_j^T J + J\bff_j \bfg_j^T=0,  \eq fgjsympl..
$$
implying that the group element $H_0(\l)$ defined in \HHlzero, \hhlzero
satisfies
$$
H_0^T(\l) J H_0(\l) = J,  \eq Hzerosympl..
$$
and hence also is in $\grSp(2s)$. It then follows that the dressed wave 
function $\Psi(\l)$ obtained by solution of the associated Riemann--Hilbert
problem also takes values in $\grSp(2s)$, and the residue matrices $N_j$
entering in the definition of the operator $\DD_\l$ have values in
$\grsp(2s)$.

   More explicitly, assuming the $(\bff_j,\ \bfg_j)$'s all have  rank $p$, 
eq. \fgjsympl implies that there exist symmetric, invertible matrices
$s_j\in \grGl(p)$ such that
$$
\bff_j = J \bfg_j s_j,  \quad s_j^T = s_j, \quad j=1, \dots s.  \eq fgsj..
$$
The orthogonality conditions \fgjnull then reduce to the symplectic isotropy
conditions
$$
\bfg_j^T J \bfg_k =0, \quad j=1, \dots s.  \eq isotr..
$$
Since the $s_j$'s are invertible and symmetric, we may take their square 
roots $s_j^{1\over 2}$ (determined up to a $\bfZ_2^r$ ambiguity) and, without
changing the content of the Riemann--Hilbert problem, redefine the $\bff_j$'s
and $\bfg_j$'s by absorbing these square roots through the substitutions
$$
\bff_j \lra \bff_j s_j^{1\over 2}, \quad \bfg_j \lra \bfg_j s_j^{-{1\over 2}}.
\eq fgjredef..
$$
This just amounts to setting the $s_j$'s equal to the identity element, 
reducing \fgsj simply to
$$
\bff_j = J \bfg_j.   \eq fgnosj..
$$
Since the solution of the Riemann--Hilbert problem preserves the
symplectic reduction, it follows from eqs. \FPsi, \GPsi and \FGjdef that the
matrices $(\bfF_j, \ \bfG_j)$ satisfy  the relations
$$
\bfF_{j} =(-1)^j  J \bfG_{j}.  \eq FGjsympl..
$$
This means that there exist pairs of rectangular $p \times s $ matrices
$(\bfQ_j, \ \bfP_j)$ such that the $\bfF_j$'s and $\bfG_j$'s may be 
expressed 
$$
\bfF_j^T = e^{i\pi j\over 2}\left(\bfP_j ,\ - \bfQ_j \right), \quad
\bfG_j^T = (-1)^j e^{i\pi j\over 2} \left(\bfQ_j ,\
\bfP_j\right), \quad j=1 \dots n.  \eq PQj..
$$
The residue matrices $N_j$ therefore have the reduced block form
$$
N_j  = - \pmatrix { \bfP_j^T \bfQ_j & \bfP_j^T \bfP_j \cr
-\bfQ_j^T \bfQ_j & -\bfQ_j^T \bfP_j},  \eq Njsympl..
$$
showing explicitly that they all belong to $\grsp(2s)$.

   The reduced form of eqs. \Fjk-- \Gjj then becomes
$$
\eqalignno{
{\di \bfQ_j \over \di \a_k} &=
 {(\bfP_j\bfQ_k^T -\bfQ_j\bfP_k^T)\bfQ_k\over \a_j - \a_k},
\quad  j\neq k,  \eqn Qjk.a. \cr
{\di \bfP_j \over \di \a_k} &=
 {(\bfP_j\bfQ_k^T -\bfQ_j\bfP_k^T)\bfP_k\over \a_j - \a_k},
\quad  j\neq k,  \eqn Pjk.b. \cr
{\di \bfQ_j \over \di \a_j} &= -\bfQ_j B_1 -\sum_{k=1 \atop k\neq j}^n
  {(\bfP_j\bfQ_k^T -\bfQ_j\bfP_k^T)\bfQ_k\over \a_j - \a_k},
  \eqn Qjj.c. \cr
{\di \bfP_j \over \di \a_k} &=  \bfP_j B_1 - \sum_{k=1 \atop k\neq j}^n
 {(\bfP_j\bfQ_k^T -\bfQ_j\bfP_k^T)\bfP_k\over \a_j - \a_k},
 \eqn Pjj.d. \cr}
$$
\nextnumber
In particular, for $p=s=1$, $r=2$, these are just the equations of 
\cite{JMMS}, Theorem 7.5, determining the $n$--particle correlation functions
for an impenetrable bosonic gas (or the spectral distribution function for
random unitary matrices in the double scaling limit \cite{TW1}).

   Similarly, the reduced form of \Fja and \Gja is given by
$$
\eqalignno{{\di \bfQ_j\over \di \b_a} = &  -\a_j\bfQ_j E_a
+ \bfQ_j\sum_{b=1 \atop b\neq a}^s {E_a N_{PQ} E_b + 
E_b N_{PQ}E_a \over \b_a -\b_b}\cr
&- \bfP_j \sum_{b=1}^s {E_a N_{QQ} E_b + E_b N_{QQ}E_a \over \b_a +\b_b}
\eqn Qja.a. \cr
{\di \bfP_j\over \di \b_a} = &  \a_j\bfP_jE_a - 
\bfP_j \sum_{b=1 \atop b\neq a}^s {E_a N_{QP} E_b +
 E_b N_{QP}E_a \over \b_a -\b_b}\cr &+ \bfQ_j \sum_{b=1}^s {E_a N_{PP} E_b +
E_b N_{PP}E_a \over \b_a +\b_b} \eqn Pja.b. \cr
a =&1, \dots s, \quad j=1, \dots n,}
$$
where
$$
N_{QQ}:= \sum_{j=1}^n \bfQ_j^T \bfQ_j, \quad
N_{PP}:= \sum_{j=1}^n \bfP_j^T \bfP_j, \quad
N_{QP}:= \sum_{j=1}^n \bfQ_j^T \bfP_j, \quad
N_{PQ}:= \sum_{j=1}^n \bfP_j^T \bfQ_j,   \eq NPQ..
$$
and $E_a$ is the elementary $s\times s$ matrix with $1$ in the $aa$ position.

Other discrete reductions of the generic Riemann--Hilbert data and the
corresponding isomonodromic deformation families may similarly be derived 
along the lines indicated in \cite{H}.

\medskip
\Subtitle {4b. Higher order poles}

\nobreak
Using arguments similar to the Zakharov-Shabat dressing method (see e.g.
\cite{NZMP}), it is possible to generalize the above considerations to a
broader class of isomonodromic deformation equations associated to matrix 
Riemann--Hilbert problems of a similar nature, by allowing higher order 
poles at one or more of the singular points. For simplicity, we restrict 
attention to the case where there is just one irregular singular point of 
arbitrary index located at $\infty$. Extension to the case of any number 
of irregular singular points is quite straightforward.

Let $\Psi_0(\l, \bft)\in \grGl(r)$ be an isomonodromic ``vacuum'' solution,
satisfying
$$
\eqalignno{
{\di \Psi_0\over \di \l} =& U_0(\l, \bft)\Psi_0,  \eqn Psivacl.a.\cr
{\di \Psi_0\over \di t_a} =& V^0_a(\l, \bft)\Psi_0, \quad a=1 \dots k
\eqn Psivacti.b.}
$$
where $\{U_0(\l, \bft),\ V^0_i(\l, \bft)\}_{i=1\dots k}$ are $r\times r$ 
matrix valued polynomials in $\l$ whose coefficients are functions of the
$k$--component vector $\bft=(t_1, \dots t_k)$. The compatibility conditions 
for \Psivacl, \Psivacti imply that the generalized monodromy data for the
``vacuum'' operator
$$
\DD^0_\l := {\di \over \di \l} -U_0(\l, \bft)  \eq..
$$
are invariant under the deformations induced by changes of the parameters
$(t_1, \dots t_k)$ satisfying these equations. For example, we could consider
the simplest case, when the matrices $V_a^0(\l)$ are independent of the
parameters $(t_1, \dots t_k)$, and commute amongst themselves
$$
\left[ V_a^0(\l), V_b^0(\l) \right] = 0, \quad a,b = 1, \dots k.  
\eq Vabcomm..
$$
The vacuum wave function then has trivial monodromy and can  be normalized as
$$
\Psi_0(\l, \bft) = \exp\left(\sum_{a=1}^k t_a V_a^0(\l)\right),  
\eq Psiabel..
$$
so that its values form an abelian group. More general cases, in which the
vacuum monodromy is not necessarily trivial may also be included (cf.
\cite{TW2-3}).

As in section 2, choose  a set $\{\bff_j, \bfg_j\}_{j=1 \cdots m}$ of fixed 
$r \times p$ rectangular matrices satisfying \fgjnull, and let $\th_j(\l)$
denote the characteristic function along the curve segment $\G_j$. Define
$$
\eqalignno{
\bff(\l, \bft)&:=   \Psi_0(\l, \bft)\sum_{j=1}^m \bff_j \th_j(\l),
 \eqn flt.a.  \cr
\bfg(\l, \bft)&:= \left(\Psi_0^T(\l, \bft)\right)^{-1}\sum_{j=1}^m \bfg_j
\th_j(\l),
\eqn glt.b. }
$$
as in \fl, \gl, but with the exponential vacuum wave function $\Psi_0(\l)$
replaced by the vacuum solution $\Psi_0(\l, \bft)$, evaluated along the
curve segments $\G_j$ (where, in the event that there is nontrivial vacuum
monodromy, a specific branch is taken).  Now define $H_0(\l)$ again as in
\HHlzero and $H(\l,\bft)$ as in \Hl and \HHl, but with the functions
$\bff(\l)$, $\bfg(\l)$  of eqs. \fl, \gl replaced by $\bff(\l, \bft)$, 
$\bfg(\l, \bft)$ and $\Psi_0(\l)$ replaced by $\Psi_0(\l, \bft)$. Let
$\chi(\l)$ again be a solution to the matrix Riemann--Hilbert problem as
defined in \MRHasympt, \MRHdisc, with the appropriate substitution of
$H(\l,\bft)$ for $H(\l)$. Then, just as in sections 2a, 2b, by considering
the local structure near the singularities and the asymptotic behaviour at
$\infty$, it is easily shown that 
$$
\Psi(\l):= \chi(\l) \Psi_0(\l, \bft)  \eq..
$$
satisfies the equations
$$
\eqalignno{
{\di \Psi\over \di \l} - U(\l, \bft)\Psi -  
\sum_{j=1}^n{N_j\over \l -\a_j}\Psi
 &=0, \eqn Psil.a.\cr
{\di \Psi\over \di t_a} - V_a(\l, \bft)\Psi &=0, \quad a=1 \dots k,
\eqn Psita.b.\cr
{\di \Psi\over \di \a_j} + {N_j\over \l -\a_j}\Psi &= 0,
\quad j=1 \dots n,  \eqn Psiaj.c. }
$$
\nextnumber
where, as before, $N_j$, $\bfF_j$, $\bfG_j$ are again given by \Nj, \FGjdef,
while  $$
\eqalignno{
U(\l, \bft) &= \left(\chi(\l) U_0(\l, \bft) \chi^{-1}(\l)\right)_+ ,
\eqn.a.\cr
V_a(\l, \bft) &= \left(\chi(\l) V^0_a(\l, \bft) \chi^{-1}(\l)\right)_+ ,
\quad a=1, \dots k,\eqn.b.}
$$
\nextnumber
where $(\ )_+$ denotes projection to the polynomial part in $\l$ 
(cf. \cite{NZMP}). It follows that the operators
$$
\eqalignno{
\DD_\l &:= {\di \over \di \l} - U(\l, \bft) - \sum_{j=1}^n{N_j\over \l -\a_j},
\eqn.a.
\cr
\DD_{t_a}&:= {\di \over \di \t_a}- V_a(\l, \bft), \quad a=1 \dots k, \eqn.b.
\cr
\DD_{\a_j}&:= {\di \over \di \a_j} +{N_j\over \l -\a_j},
\quad j=1 \dots n \eqn.c.}
$$
all commute, and the generalized monodromy data for $\DD_\l$ is invariant 
under the deformations induced by changes in the parameters $\{\a_j\}_{j=1
\dots n}$, $\{t_a\}_{a=1 \dots k}$.

The $\t$ function associated with these deformations may once again be shown 
to be the Fredholm determinant $\det(\bfI-\bfK)$ of the integral operator
$\bfK$ defined as in \FredK, \Kdef, with the appropriate substitutions for
$\bff(\l)$, $\bfg(\l)$. The coefficients $\{H_j\}_{j=1 \dots n}$
$\{K_a\}_{a=1 \dots k}$ of the differential form
$$
d\ln \det(\bfI -\bfK)= \sum_{i=1}^nH_i d\a_i
+ \sum_{a=1}^k K_a dt_a \eq..
$$
may again be viewed as commuting nonautonomous Hamiltonians whose Hamiltonian
equations determine the deformation conditions corresponding to the
commutativity of the operators $\DD_\l$, $\DD_{t_a}$, $\DD_{\a_j}$. For the
specific case \Psiabel, where the vacuum wave functions form an abelian 
group, the same interpretation may be given to the corresponding
$\tau$--function as in Theorem 2.6. More generally, if we drop the
requirement that \Wgammat define an abelian group action on the Hilbert space
Grassmannian $\HH^r$, and simply view the maps determined by \Wgammat \amap,
\bmap, \dmap as defining a set of integral curves for a compatible
nonautonomous system generated by the solution $\Psi_0(\l, \bft)$ of the
associated vacuum system, the same geometrical interpretion in terms of
determinants of projection operators is still valid. The resulting
differential operators $\DD_\l$, $\wt{\DD_z}$ may also be expressed in terms
of a Hamiltonian quotienting procedure through formulae similar to \NNintr,
\MMintr, and we may thereby again associate a dual system and corresponding
dual integral operator $\wt{\bfK}$, as was done in section 3. Details of
this, as well as generalizations involving multiple higher order poles will
be given elsewhere.

\bigskip\bigskip \noindent{\it Acknowledgements.}
This research was supported in part by the Natural Sciences and
Engineering Research Council of Canada, the Fonds FCAR du Qu\'ebec and 
the National Science Foundation, grant No. DMS-9501559.
\bigskip \bigskip
\goodbreak
\references

DIZ& Deift, P.~A., Its, A.~R., and Zhou, X., ``A Riemann-Hilbert Approach
to Asymptotic Problems Arising in the Theory of Random Matrix Models, 
And Also in the Theory of Integrable Statistical Mechanics''
{\it Ann. of Math.} (1997) (in press).

H& Harnad, J., ``Dual Isomonodromic Deformations and Moment Maps to Loop
Algebras'' {\it Commun. Math. Phys.} {\bf 166}, 337--365 (1994).

HTW&  Harnad, J., Tracy,  C.~A.,  and  Widom, H.,
``Hamiltonian Structure of Equations Appearing in Random Matrices'' , in:
{\it  Low Dimensional Topology and Quantum Field Theory}, ed. H. Osborn,
(Plenum,  New York  1993), pp. 231--245.

IIKS& Its, A.~R., Izergin, A.~G., Korepin, V.~E., and Slavnov,
N.~A., ``Differential Equations for Quantum Correlation Functions,''
{\it Int.~J.~Mod.~Phys.\/}  {\bf
B4}, 1003--1037 (1990).

IIKV& Its, A.~R., Izergin, A.~G., Korepin, V.~E.,  and Varzugin,
G.~G., ``Large Time and Distance Asymptotics of Field Correlation Function of
Impenetrable Bosons at Finite Temperature,''  {\it Physica\/}  {\bf 54D},
351--395 (1992).

JMMS& Jimbo, M., Miwa, T., M{\^ o}ri, Y., and Sato, M., ``Density
Matrix of an Impenetrable Bose Gas and the Fifth Painlev\'e Transcendent'',
{\it Physica\/} {\bf 1D}, 80--158 (1980).

JMU& Jimbo, M., Miwa, T., and  Ueno, K., ``Monodromy Preserving
Deformation of Linear Ordinary Differential Equations with Rational
Coefficients I.,''  {\it Physica\/} {\bf 2D}, 306--352 (1981).

KBI&  Korepin, V.~E.,  Bogolyubov, N.~M., and  Izergin, A.~G.,
{\it Quantum Inverse Scattering \break Method  and Correlation Functions},
Cambridge Monographs on Mathematical \break
Physics, Cambridge (1993).

M& Mehta, M.~L., {\it Random Matrices}, 2nd edition (Academic, San Diego, 
1991).

NZMP& Novikov, S.~P., Zakharov, V.~E., Manakov, S.~V., Pitaevski, L.~V.,
 `` Soliton Theory: The Inverse Scattering Method'' {\it Plenum, New York}
 (1984).

P& Palmer, John, ``Deformation Analysis of matrix models'', {\it
Physica} {\bf D 78}, 166--185 (1995).

Sa& Sato, M., ``Soliton equations as dynamical system on infinite dimensional
Grassmann manifolds'', {\it  RIMS Kokyuroku} {\bf 439}, 30 (1981).

SW& Segal, G. and Wilson, G., ``Loop Groups and Equations of KdV Type'',
{\it  Publications Math., I.H.E.S.} {\bf 61}, 5--65 (1985).

TW1& Tracy, C.~A., and Widom, H., ``Introduction to Random
Matrices,''   in: {\it 	Geometric and Quantum Methods in Integrable Systems.}
 Springer Lecture Notes in Physics {\bf	424}, pp. 103--130
(ed. G. F. Helminck, Springer--Verlag, N.Y.,  Heidelberg,  1993).

TW2& Tracy, C.~A., and Widom, H., ``Fredholm determinants, differential
equations and matrix models''   {\it Commun. Math. Phys.} {\bf 163}, 33--72
(1994).

TW3& Tracy, C.~A., and Widom, H., ``Systems of partial differetial equations
for operator determinants''   {\it Oper. Th.: Adv. Appl.} {\bf 78}, 381--388
(1995).

W& Wilson, G., ``Habillage et Fonction $\tau$'', {\it C.R. Acad. Sc.,
Paris},  {\it 299}, 587--590 (1984).

WMTB& Wu, T.~T., McCoy, B.~M., Tracy, C.~A., and  Barouch, E., ``Spin-Spin
Correlation Functions for the Two-Dimensional Ising Model: Exact Theory in 
the Scaling Region,'' {\it Phys.\ Rev.}  {\bf B13}, 316--374 (1976).

\endreferences

\end